\documentclass[
  journal=largetwo,
  manuscript=article-type,
  year=2025,
]{cup-journal}

\usepackage{amsmath}
\usepackage[nopatch]{microtype}
\usepackage{booktabs}
\usepackage{hyperref}
\usepackage{tabularx}
\usepackage{soul}
\usepackage{float}
\usepackage{subcaption}
\title{First Use of GPS Satellites for Beam Calibration of Radio Dish Telescopes}
\DeclareUnicodeCharacter{2009}{\,} 

\author{S. Berger}
\affiliation{School of Physics, University of Melbourne, Parkville, VIC 3010, Australia}
\alsoaffiliation{Department of Physics, McGill University, Montréal, QC, H3A 2T8, Canada}
\alsoaffiliation{Trottier Space Institute, McGill University, Montréal, QC, H3A 2T8, Canada}
\email[Sabrina Berger]{berger@student.unimelb.edu.au}

\author{A. Lasinski}
\affiliation{Department of Physics, McGill University, Montréal, QC, H3A 2T8, Canada}
\alsoaffiliation{Trottier Space Institute, McGill University, Montréal, QC, H3A 2T8, Canada}
\alsoaffiliation{Department of Astronomy and Astrophysics, University of Toronto, Toronto, ON M5S
3H4, Canada}

\author{V. MacKay}
\affiliation{MIT Kavli Institute for Astrophysics and Space Research, 77 Massachusetts Ave., Cambridge, MA 02139, USA}
\author{E. Egan}
\affiliation{Department of Physics, McGill University, Montréal, QC, H3A 2T8, Canada}
\alsoaffiliation{Trottier Space Institute, McGill University, Montréal, QC, H3A 2T8, Canada}

\author{D. Wulf}
\affiliation{Department of Physics, McGill University, Montréal, QC, H3A 2T8, Canada}
\alsoaffiliation{Trottier Space Institute, McGill University, Montréal, QC, H3A 2T8, Canada}

\author{A. Chokshi}
\affiliation{Department of Physics, McGill University, Montréal, QC, H3A 2T8, Canada}
\alsoaffiliation{Trottier Space Institute, McGill University, Montréal, QC, H3A 2T8, Canada}

\author{J. Sievers}
\affiliation{Department of Physics, McGill University, Montréal, QC, H3A 2T8, Canada}
\alsoaffiliation{Trottier Space Institute, McGill University, Montréal, QC, H3A 2T8, Canada}


\begin{document}

\begin{abstract}
We present results from the first application of the Global Navigation Satellite System (GNSS; e.g., the Global Positioning System, GPS) for radio beam calibration using a commercial GNSS receiver with the Deep Dish Development Array (D3A) at the Dominion Radio Astrophysical Observatory (DRAO). Several GNSS satellites pass through the main and sidelobes of the beam each day, enabling efficient mapping of the 2D beam structure. Due to the high SNR and abundance of GNSS satellites, we find evidence that GNSS can probe several sidelobes of the beam through repeatable measurements of the beam over several days. Over three days of measurements, the smallest observed difference in the primary beam’s main lobe was 0.56 dB-Hz. We also compare our results in the sidelobes to simulations and find rough agreement in shape. When scaling the observations and simulations to match in the main lobe power levels, we find deviations in at least one of the first few nulls of approximately 5 dB or less. There is saturation in the main lobe for most satellites, which can likely be mitigated by better attenuation before the receiver input.  We compare our work to other satellite systems that have been successful and are likely complementary to this technique. However, GNSS offers key advantages, including continuous transmission, broader frequency coverage relevant to CHORD, SKA-mid, and the DSA-2000, as well as more frequent satellite passes, making it a promising calibration method. These results also motivate further development of this technique for radio astronomy applications.

\end{abstract}

\section{Introduction}
\label{sec:intro}

The primary challenge in detecting cosmological 21\,cm emission lies in the brightness of astrophysical foregrounds; effective removal depends on an exquisite understanding of the instrument’s response, including detailed beam characterization. The Canadian Hydrogen Observatory and Radio-transient Detector (CHORD) is a new radio telescope with 21\,cm intensity mapping among its primary science goals. CHORD will be built at the Dominion Radio Astrophysical Observatory (DRAO) near Penticton, British Columbia. It will operate as a drift-scan instrument with 512 dishes that can be manually repointed to different declinations, building on the success of its predecessor, the Canadian Hydrogen Intensity Mapping Experiment (CHIME), also a drift-scan telescope. Drift-scan instruments like CHIME and CHORD have fixed pointing geometries that cannot be adjusted in real time, making it challenging to obtain complete beam measurements across the full sky. CHORD shares many of CHIME's science goals, notably low-redshift 21 cm intensity mapping and fast radio burst (FRB) detection and localization with very-long-baseline interferometry (VLBI), but has significantly boosted sensitivity due to major instrumental innovations---in particular, a threefold bandwidth expansion from 400--800 MHz to 300--1500 MHz.  Additionally, CHORD's dishes are designed for sub-millimeter surface precision (1/1000 of a wavelength at 300\,MHz), and advances in composite fabrication show that this specification can be met or exceeded \citep{metrology}. Dish surface precision is thus not expected to limit CHORD's redundant calibration scheme \citep{red_cal}; instead, the primary bottleneck will be measuring and correcting differential beam variations between antennas, underscoring the need for precise beam mapping—an issue CHIME has already highlighted in its effort to detect the Baryon Acoustic Oscillations (BAO) at low redshifts \citep{chime_paper_latest}.

Traditional beam mapping with steerable telescopes uses bright astrophysical sources raster-scanned across the field of view; such techniques are unavailable to drift-scan arrays. Instead, for CHIME and its pathfinder, holography with the adjacent 26-m John A. Galt telescope has been used \citep{Berger_2016,amiri2024holographicbeammeasurementscanadian}, providing detailed East-West coverage but sparse sampling in declination, and requiring long integrations due to low-signal-to-noise (SNR) sources. CHIME has also employed solar drift scans \citep{beam_map}, though those rely on minimal solar variability. More generally, few astrophysical sources are bright enough to probe sidelobes, and their fixed trajectories cover only a small solid angle. These limitations are most salient for ``large-$N$, small-$D$'' interferometers, whose many small elements provide wide primary fields of view, making beam-mapping methods with broad angular coverage essential.

Artificial calibration sources have recently expanded the toolkit. Drone-based methods allow precise control of both the transmitted signal and flight path, and have been used to map full beams, including the first drone-based beam measurement of a cylindrical interferometer with CHIME \citep{Chang_2015,bolli_2016,2024arXiv240700856H,2024arXiv240704848T}. Holographic, solar, and drone-based measurements have shown good agreement \citep{amiri2024holographicbeammeasurementscanadian}, particularly in identifying the main lobe and sidelobes. However, drones are limited by far-field requirements ($>$300\,m for a 6\,m dish at 21\,cm). Satellites, by contrast, are always in the far field for instruments like CHORD. ORBCOMM satellites, for example, have been used to map the Murchison Widefield Array at $\sim$100\,MHz \citep{neben,Line_2018,Chokshi_2021}. Among potential satellite calibrators, the Global Navigation Satellite System (GNSS) is a particularly promising candidate. GNSS encompasses more than 100 high-SNR satellites in medium Earth orbit ($\sim$20,000\,km), with operating frequencies (1100--1600\,MHz) that overlap with CHORD’s band. Previous work demonstrated feasibility with the EMBRACE SKA-prototype array using GPS satellites at 1227.6\,MHz, mapping beams out to 45$^\circ$ from boresight \citep{2009wska.confE..42O,Torchinsky_2016}.

GNSS is also being explored as a phase calibrator for VLBI, providing total electron content (TEC) measurements at each site. This is particularly valuable for real-time FRB localization, where no traditional calibrator is available at the time of the burst. Achieving useful precision requires careful receiver characterization \citep{coster_accuracy}, but mapping efforts are approaching the $\sim$0.01 TECu accuracy needed for fringe fitting. Carrier-phase methods promise even higher precision, since GNSS and VLBI share the same atomic clocks. For example, NIST has demonstrated frequency comparisons with $10^{-15}$ precision over 24 hours \citep{precision_GNSS_carrier}. Ongoing work by NIST and related efforts (e.g., Common View GPS Time Transfer\footnote{\url{https://www.nist.gov/pml/time-and-frequency-division/time-services/common-view-gps-time-transfer}}) highlights the potential of GNSS as a prior for initial VLBI phase calibration.

To test this approach for CHORD, we use the Deep Dish Development Array (D3A), a prototype of three 6\,m dishes operating from 300--1600 MHz and built to test CHORD's new technologies. Using an off-the-shelf GNSS receiver, we present the first beam map of a radio dish measured with GNSS. Our analysis shows sensitivity into the sidelobes with daily repeatability, demonstrating GNSS as a valuable complement to existing calibration methods. With the abundance of satellites, GNSS offers large coverage at CHORD frequencies and may also enable lossless recovery of low-$z$ 21\,cm measurements \citep{Harper_2018}.

In this paper, we provide a proof of concept for GNSS-based beam mapping with CHORD. Section~\ref{sec:experiment} describes the experimental setup, simulations, GNSS mapping technique, and receiver. Results are presented in Section~\ref{sec:results} and compared to simulations. Section~\ref{sec:discussion} discusses how this approach complements other calibration methods and possible developments, and Section~\ref{sec:conclusion} summarises our findings.

\section{Experimental Setup}
We provide a brief overview of the radio telescope beam, the D3A instrument, GNSS, and the commercial receiver (Septentrio Mosaic X-5 Receiver). We direct the reader to \citet{2009tra..book.....W} and \citet{misra2011global} for more extensive overviews of the radio telescope beam and GNSS, respectively.
\label{sec:experiment}

\subsection{Physical Radio Telescope Beams}
\label{sec:radio_beams}
The antenna beam describes the sensitivity to different parts of the sky as a function of the angle of incoming radiation. Unless otherwise stated, the ``beam'' will henceforth refer to the power pattern, which is equal to the square of the electric field strength. It can be obtained by taking the Fourier transform of the aperture illumination pattern by the feed; while this can sometimes be derived analytically or computed numerically, it is not always straightforward with imperfect or non-uniform illumination, and precision calibration of an antenna generally requires experimental beam mapping. In this paper, we aim to map the beam of the individual antennas forming the D3A, referred to as their \textit{primary beams}---as opposed to the \textit{synthesized beam} of the interferometer---which we will denote $P(\theta, \phi, \lambda)$, where $\theta$ is the elevation, $\phi$ is the azimuth, and $\lambda$ is the observing wavelength as described in \citet{2009tra..book.....W}. It is common to present the peak-normalised beam, defined as
\begin{equation}
\label{eq:beam_pattern}
    P_\mathrm{norm}(\theta, \phi, \lambda) = \frac{1}{P_\mathrm{max}}P(\theta, \phi, \lambda),
\end{equation}
where $P_\mathrm{max}$ is the (frequency-dependent) maximum of $P(\theta,\phi,\lambda)$ over all values of $\theta$ and $\phi$. While most of a beam’s power is concentrated in its \textit{main lobe}---the region between $\theta = 0^\circ$ and the first null---a non-negligible fraction also resides in the \textit{sidelobes} (from the first null to $\theta = 90^\circ$) and in the \textit{backlobes} ($\theta > 90^\circ$). For example, in the idealized case of a uniformly illuminated circular aperture, the beam reduces to an Airy disk, which is the Fourier transform of a disk:
\begin{equation}
    P(\theta) = P_0 \left[\frac{2 J_1 (k~r \sin\theta)}{k~r\sin\theta}\right]^2,
    \label{eq:airy}
\end{equation}
where $P_0$ is the maximum intensity of the beam, $J_1$ is a first order Bessel function of the first kind, $k=\frac{2\pi}{\lambda}$ is the wavenumber associated with the observing wavelength $\lambda$, $r$ is the radius of the aperture (3\,m for the D3A), and $\theta$ is the angle from the axis of the circular aperture (the $\phi$-dependence was left out as the Airy pattern is axisymmetric). In the idealized case for a dish with a circular aperture, the size of the main lobe is that defined by the first nulls of the corresponding Airy disk, 
\begin{equation}\theta_{\rm main~lobe} = 1.22\frac{\lambda}{d}.
\end{equation}

where $d$ is the diameter of the dish. For such a beam, the main lobe represents approximately 84\% of the power; thus, the remainder of the beam accounts for 16\% of the received power.

\subsection{Simulated D3A Beams}
\label{sec:simulated}
\begin{figure*}[t]
    \centering
    \begin{subfigure}{0.48\textwidth}
        \includegraphics[width=\textwidth]{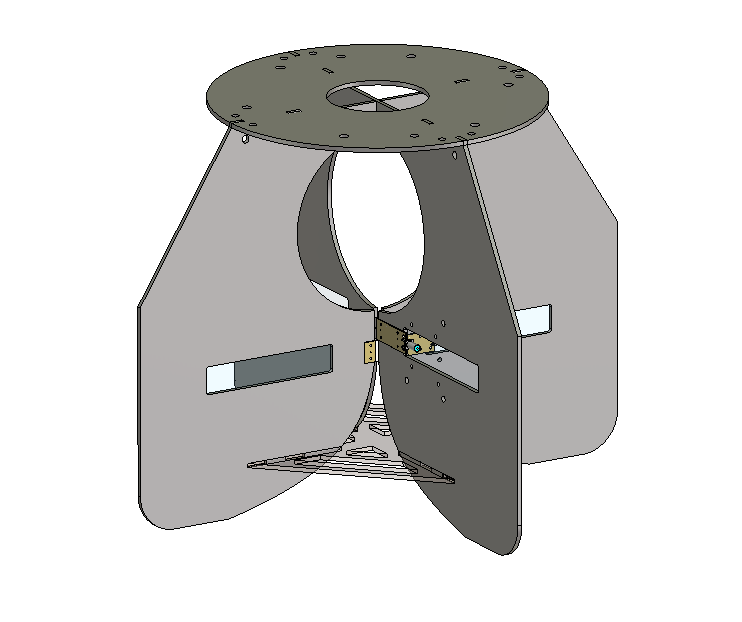}
        \caption{}
        \label{fig:feed}
    \end{subfigure}
    \hfill
    \begin{subfigure}{0.48\textwidth}
        \includegraphics[width=\textwidth]{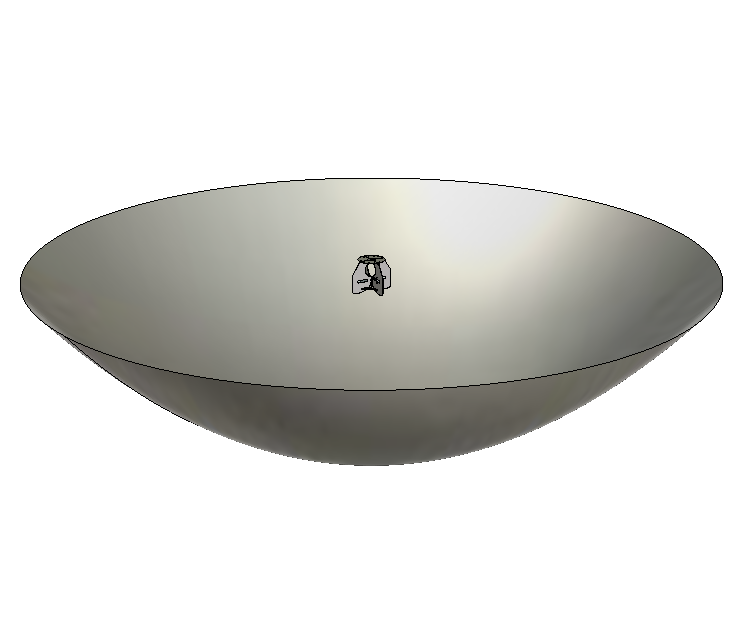}
        \caption{}
        \label{fig:feed_dish}
    \end{subfigure}
    \caption{Model of a single feed and dish as simulated in CST Microwave Studio, (a) zooming in on the feed, and (b) showing the whole model. The dish was simplified to an analytic parabola and all support structures were omitted, to enable a more efficient simulation.}
    \label{fig:cst_models}
\end{figure*}

\begin{figure*}[t]
    \centering
    \begin{subfigure}{0.48\textwidth}
        \includegraphics[width=\textwidth]{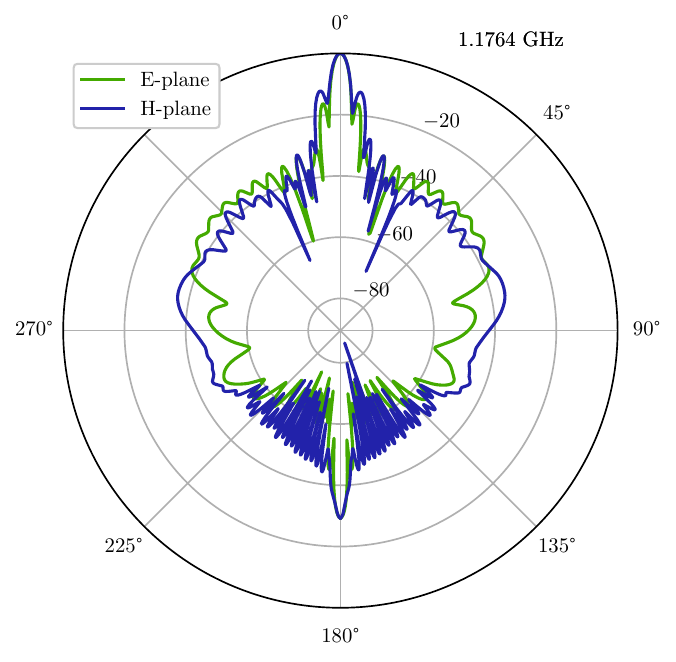}
        \caption{}
        \label{fig:1.18_sim}
    \end{subfigure}
    \hfill
    \begin{subfigure}{0.48\textwidth}
        \includegraphics[width=\textwidth]{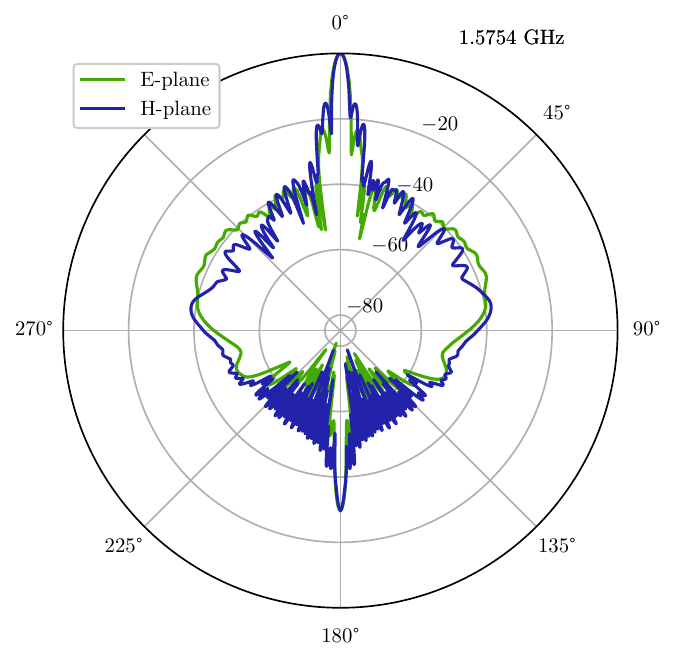}
        \caption{}
        \label{fig:1.5_sim}
    \end{subfigure}
    \caption{Peak-normalised beams in polar coordinates for the E and H planes at (a) 1.1764 GHz and (b) 1.5754 GHz, as simulated with CST Microwave Studio using the models shown in Figure~\ref{fig:cst_models}. Those frequencies correspond to GPS L5 and L1, respectively.}
    \label{fig:twopanel}
\end{figure*}

We compare our GNSS observations to those from a simulated D3A dish in the electromagnetic (EM) field solver \href{https://www.3ds.com/products/simulia/cst-studio-suite}{CST Microwave Studio}. The simulations were run in a single-dish configuration (thus excluding antenna-to-antenna coupling) using the Vivaldi feed deployed during the August 2022 measurements \citep{2023JAI....1250008M}. The model of the feed and dish can be seen in Figure \ref{fig:feed_dish}. To enable faster computation, the dish was modeled as an analytic parabola with simplified geometry: support structures and the curved rim were omitted, as was the coaxial cable running through one of the four support legs (Figure~\ref{fig:cst_models}). While the support legs for the D3A dishes are made of fiber-reinforced plastic and should have minimal effect on the beam at GNSS frequencies, the absence of the coaxial cables may introduce more noticeable differences, which we acknowledge as a limitation of this work. Including the thin coaxial cables would require a much finer mesh, and thus substantially higher computational costs, so it was excluded for the purpose of this preliminary estimate. Figures~\ref{fig:1.18_sim} and \ref{fig:1.5_sim} show the peak-normalised beam at the GPS L1 and L5 frequencies, respectively. For comparison with our measurements, we 
apply a regular grid interpolator \citep{2020SciPy-NMeth} to obtain beam values in arbitrary directions, since satellite trajectories do not align with the native simulation grid.

\subsection{The Deep Dish Development Array (D3A)}
\label{sec:D3A}

\begin{figure}
    \centering
    \subfloat[]{{\includegraphics[width=\linewidth]{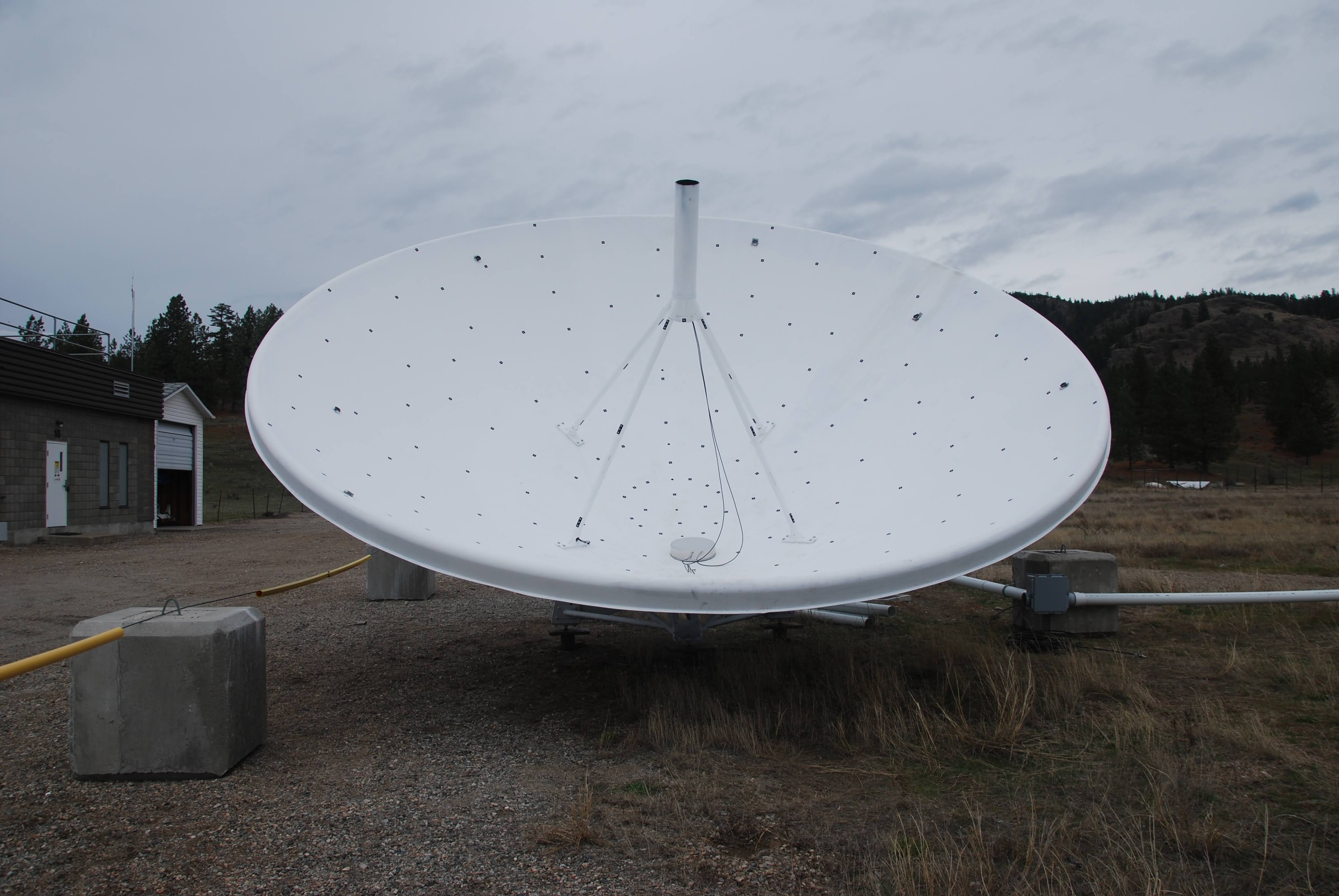} }}
    \qquad
    \subfloat[]{{\includegraphics[width=\linewidth]{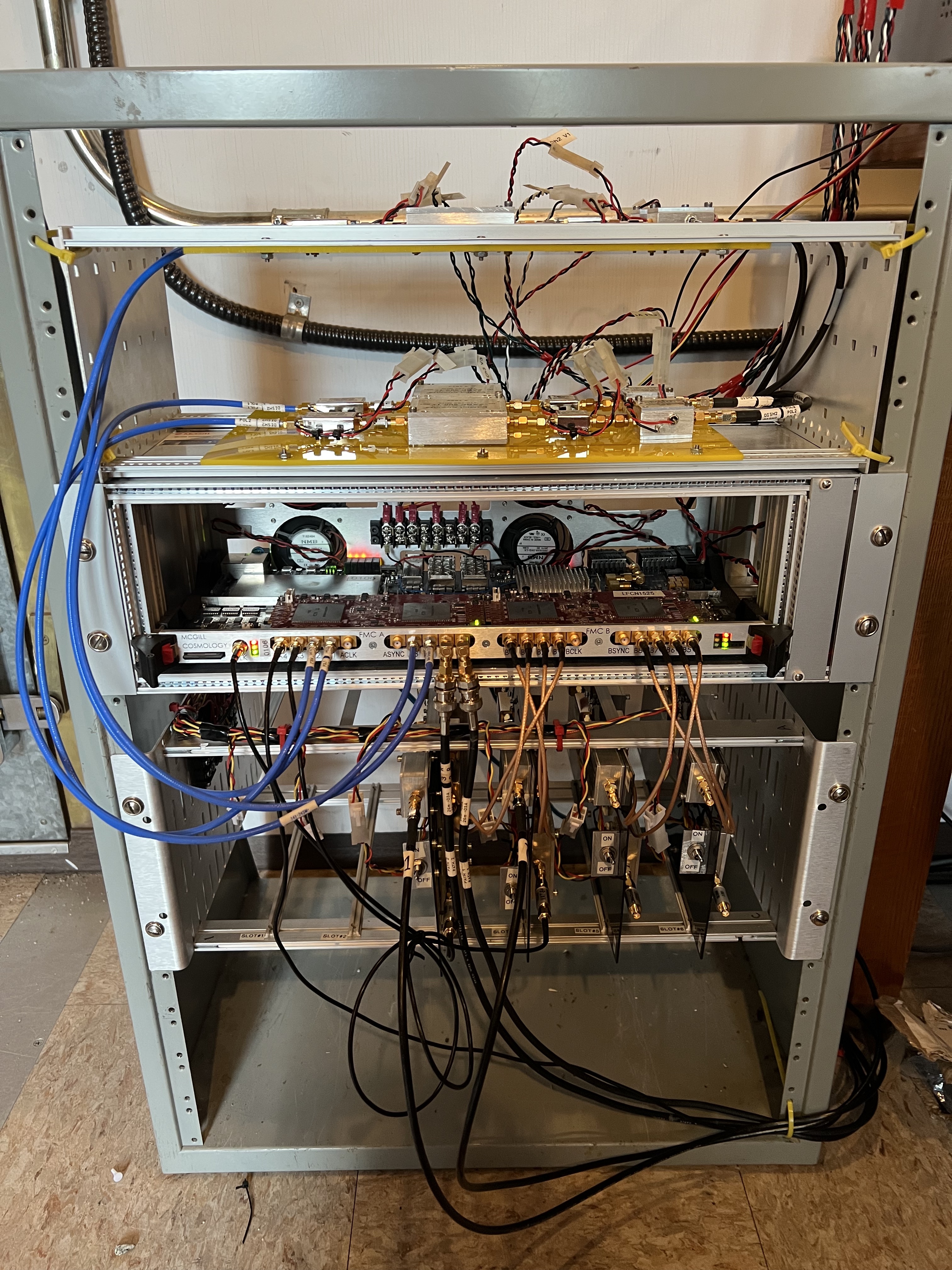} }}%
    \caption{Observation setup at the DRAO; (a) one of the three 6\,m D3A dishes---CHORD will include 512 similar dishes situated next to the main CHIME site---(b) backend of the D3A. The bottom crate slot contains the triplexer used in data collection. The middle part of the crate shows the ICE boards normally used in processing the D3A signal.\label{fig:d3a_dish_crate}}
\end{figure}

The D3A is an early prototype for CHORD, consisting of three 6\,m dishes used to test new technologies such as metrology and ultra-wideband receivers. One of the dishes is shown in the top panel of Figure~\ref{fig:d3a_dish_crate}. The D3A dishes have focal ratios of $f/d=0.25$ (CHORD dishes will be deeper, at $f/d=0.21$) and are equipped with the Vivaldi feed described in \citet{2023JAI....1250008M}. The resulting azimuthally averaged full widths at half maximum (FWHM) are of $9.9^{\circ}$ at 300\,MHz, $2.2^{\circ}$ at 900\,MHz, and $1.5^{\circ}$ at 1500\,MHz (CHORD's lowest, middle, and highest frequencies, respectively).

At the time of the measurements, D3A's signal chain was equipped to observe over the 400--1600\,MHz band, allowing it to cover all but the highest GNSS frequencies. The full electronics rack, including the RF chains, is shown in the bottom panel of Figure~\ref{fig:d3a_dish_crate}. After amplification and filtering, the RF signal is split by a prototype triplexer into three bands: 400--800\,MHz, 800--1200\,MHz, and 1200--1600\,MHz. The D3A digitization is usually performed with multiple ICE boards \citep{iceboard}, which sample the signal with 8-bit precision. However, for GNSS beam mapping, we bypass this chain and take the RF signal prior to the ICE boards. Future GNSS radiometric beam measurements will require pre-processing directly at the voltage level from the ICE boards.

The D3A power is scaled for this system, with an absolute output of -20 dBm over 400 MHz of bandwidth. We found that applying an additional 3 dB attenuation produced a signal level well matched to the input of the Septentrio receiver.

\subsection{GNSS}
\label{sec:GNSS}
GNSS refers to the global constellations of satellites used for positioning, timing, and atmospheric observations, including the Global Positioning System (GPS). These satellites transmit across 1100--1605\,MHz, overlapping with CHORD’s observing band. At any given time, tens of GNSS satellites (often 40 or more) are above horizon, leading to high chances of having one pass through a D3A dish’s primary beam and thus provide multiple measurement opportunities, at various angles. Occasionally, more than one satellite lies within a D3A dish’s primary beam's main lobes, but since their orbits are precisely known as a function of time, these moments can be flagged and taken into account in analysis---a complication we revisit in Section~\ref{sec:discussion}. Overall, the combination of high satellite density and wide range of transit angles makes GNSS particularly advantageous for beam calibration.

\begin{figure*}[h!]
    \centering
    \includegraphics[width=0.9\linewidth]{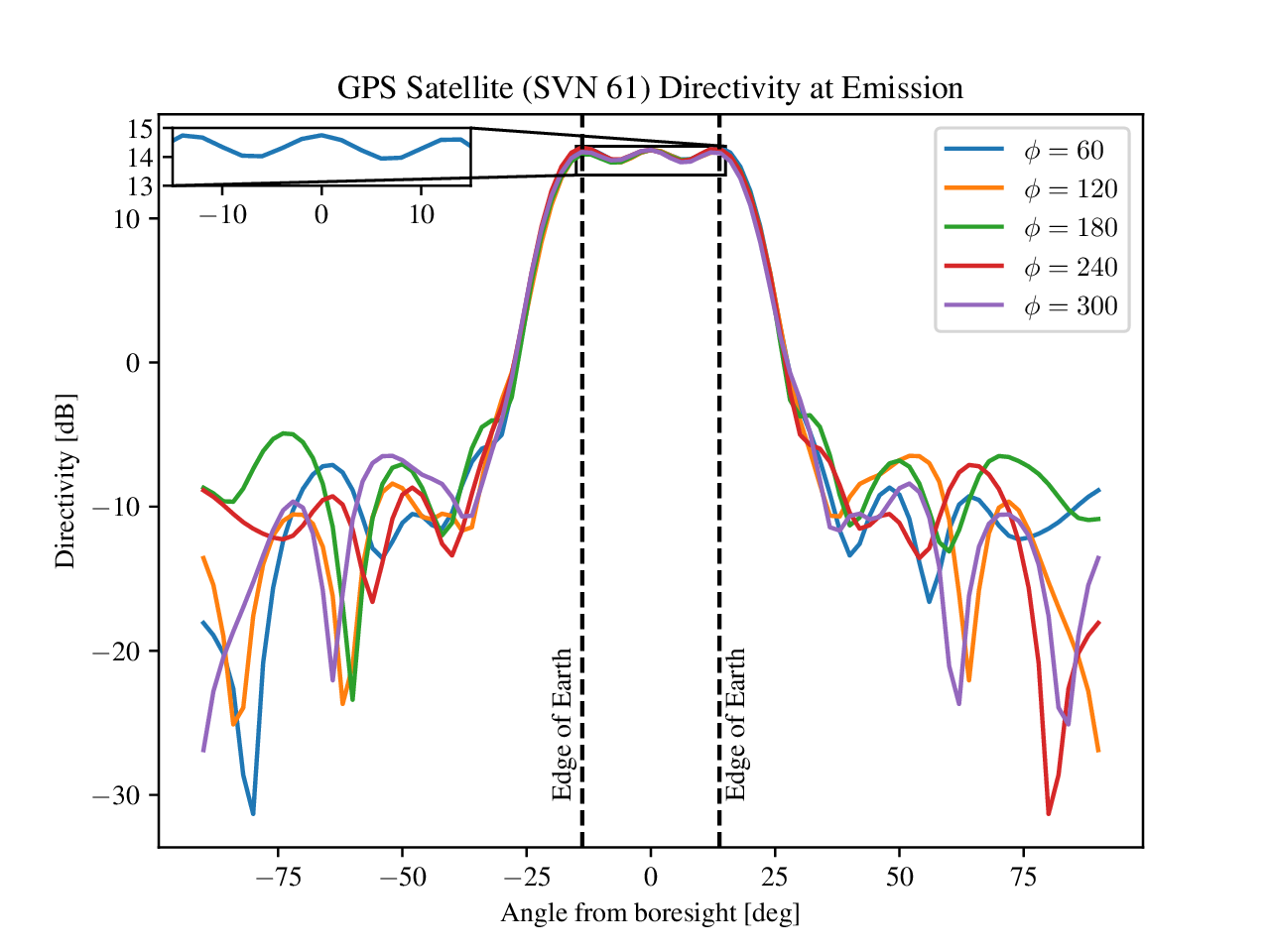}
    \caption{Plot of GPS SVN 61's directivity at emission, i.e., characterised from the on-satellite antenna panel. The zoomed in plot exemplifies the fact that there is only up to about 2 dB power loss across the Earth. The data used to make this plot were taken in 10 degree increments from the US Coast Guard Navigation Website. }
    \label{fig:gnss_beam}
\end{figure*}

The various GNSS constellations include GPS (United States), GLONASS (Russia), Galileo (European Union), and BeiDou (China). The true transmit power of GNSS satellites is not publicly available, though it is expected to remain relatively constant compared to other constellations such as ORBCOMM \citep{GNSS_antenna_characterization}. While broadcast power is undisclosed, the directivity---the ratio of intensity in a given direction to the average intensity in all directions---is known for some satellites. In particular, directivity and gain patterns have been published for GPS satellites \citep{GPS_pattern}. Figure~\ref{fig:gnss_beam} shows the measured directivity pattern of GPS Space Vehicle Number (SVN) 61 (also known as GPS IIR-13, USA-180). SVN 61 incorporated an improved antenna panel designed to increase power toward the edges of the Earth’s disk. The pattern was measured by Lockheed Martin with an uncertainty of $\pm 0.25~\rm dB$. As illustrated in Figure~\ref{fig:gnss_beam}, GNSS beams are engineered to be azimuthally symmetric and nearly constant across the full visible Earth. For this reason, we assume the received power to be constant along each satellite track. \citet{hurtado_2001} found no more than a 3 dB variation in received power across all elevation angles.

In this work, we take advantage of the fact that each GNSS satellite has a stable, nearly invariant beam profile, which allows us to perform relative beam mapping (average and 
standard deviation, see Section~\ref{sec:indiv}); absolute calibration, however, would require comparison with a well-characterised reference antenna, such as one measured in an anechoic chamber, as 
demonstrated in \cite{GNSS_antenna_characterization}. Absolute power measurements can thus be obtained by combining data from that reference antenna through a best fit, least-squares approach. A similar technique could be implemented with reference antennas adjacent to future CHORD dishes and will ultimately be required for the high-precision beam measurements needed for 21\,cm cosmology.

GPS 
places its satellites in semisynchronous orbits, completing two revolutions every sidereal day (23 hours, 56 minutes). This produces repeating ground tracks relative to the stars, making satellite passes highly predictable. As a result, only some satellites pass directly overhead at a given location, while others remain at lower elevations, constraining which can be used for observations. GPS employs six orbital planes with four satellites each, and their orbital nodes slowly regress due to Earth’s gravitational perturbations at a rate of about –0.04$^\circ$/day \citep{USDoD2008}. Although we do not account for errors from nodal regression or from satellite launches in this preliminary study, these effects should be included in future GNSS beam maps.

The GNSS signal structure \citep{8002960, Raghavan} varies with satellite type. Because of this, we focus primarily on describing GPS and the general GNSS signal characteristics. Each GPS centre frequency is a multiple of the 10.23\,MHz master GPS clock. GPS frequencies are all in the L-band and are named as follows: $\rm L5 = 1176.45 \rm MHz = 115 \times 10.23 \rm MHz$, $\rm L2 = 1227.6 \rm MHz = 120 \times 10.23 \rm MHz$, $\rm L1 = 1575.42 \rm MHz = 154 \times 10.23 \rm MHz$. The full range of GNSS frequencies is quite comprehensive, coarsely spanning a 1.1-1.6 GHz range \footnote{See a full list of GNSS frequencies (as of May, 2024) \href{https://en.gemsnav.com/news/GNSS_Frequencies_and_Signals}{here}.} 
Several satellites can transmit at the same frequency due to code division multiple access (CDMA). For GPS, this is accomplished with pseudorandom noise (PRN) codes. Each satellite has a particular PRN code corresponding to its space vehicle number which allows the GNSS receiver to detect visible satellites in a frequency channel. The GNSS satellite modulates a carrier wave with a binary PRN code with various modulation schemes. In the United States' GPS constellation, there are currently approximately 31 active satellites, each transmitting a unique PRN code. Each of the other GNSS satellites in the international array also has an associated unique PRN code. This number changes each year as new satellites are launched and decommissioned.  For GPS, there is both the Coarse/Acquisition (C/A) and military precision (P-code) with chipping rates of 1.023 Megachips/s and 10.23 Megachips/s, respectively. C/A codes repeat every millisecond after 1023 chips whereas P-codes repeat weekly. Commercial receivers can decode only C/A codes because P-codes are still encrypted by the US government.  

For the L1 GPS signal, the PRN is modulated into a GPS carrier wave through the binary phase shift key (BPSK) modulation scheme.  The binary message is encoded in the PRN code using two different phases of the carrier wave. $\theta = 0 ^{\circ}$ corresponds to a binary value of 1 and $\theta = 180 ^{\circ}$ corresponds to a binary value of 0. However, this modulation scheme is not standardized across all of GNSS or even GPS. For example, BeiDou's individual signals vary in modulation schemes from BPSK at 1176.45\,MHz to Quadrature Multiplexed Binary Offset Carrier (QMBOC) at 1575.42\,MHz. Navigation data are modulated at 50 bits per second and include satellite ephemeris data and health status. As seen in \citep{misra2011global}, the L1 GPS signal can be described as follows where each variable shown in Equation~\ref{eq:sig} is described briefly in Table \ref{table:sig},
\begin{align}\label{eq:sig}
    s(t) = \sqrt{2P_{c}} N(t) PRN(t) \cos(2 \pi ft + \theta) &+ \nonumber \\ \sqrt{2P_{p}} N(t) P_{code}(t) \sin(2 \pi ft + \theta).
\end{align}

In GNSS, received signal strength is usually expressed as the 
carrier-to-noise density ratio, $C/N_0$ in $\mathrm{dB\!-\!Hz}$. 
This quantity represents the ratio of the carrier power to the noise 
power per unit bandwidth, and is independent of the receiver’s front-end 
bandwidth.  It is sometimes convenient to convert $C/N_0$ to a conventional 
signal-to-noise ratio (SNR). The relation is
\begin{equation}
\mathrm{SNR_{dB}} \;=\; (C/N_0)_{\mathrm{dB\!-\!Hz}} \;-\; 10 \log_{10}(B),
\end{equation}
where $B$ is the effective bandwidth of the receiver. For the L1 C/A signal, the spreading bandwidth is about 2\,MHz, since the 1.023\,MHz chipping rate produces a main lobe that extends roughly $\pm 1.023$\,MHz from the carrier. Using a receiver bandwidth much wider than this simply integrates additional noise without capturing more signal, thereby lowering the effective SNR. This explains why the choice of $B$ matters: the signal is fully contained within the $\sim$2\,MHz spread, while the noise power scales directly with the selected receiver bandwidth. In contrast, the P(Y) code has a 10.23\,MHz chipping rate, corresponding to a spread of $\sim$20\,MHz. For this reason, $C/N_0$ is the preferred metric in GNSS: it is independent of receiver bandwidth, directly reflects signal quality, and avoids ambiguity in SNR definitions, which depend on both bandwidth 
and measurement point in the receiver chain.

\begin{table}[h!]
    \begin{tabularx}{0.7\textwidth}{c|X}
        \textbf{Variable}  & \textbf{Brief Description}  \\
        \hline
        $P_{c}$ & Amplitude of the C/A code \\
        $P_{p}$ & Amplitude of the P-code \\
        $N(t)$ & Navigation data \\
        $PRN(t)$ & C/A code as pseudorandom noise code \\
        $f$ & Frequency of signal (L1 here) \\
        $P_{code}(t)$ & Military P-code \\
        $\sin(2 \pi ft + \theta)$ & RF carrier
    \end{tabularx}
    \caption{Overview of variables shown in Equation~\ref{eq:sig} which describes the signal as a function of time.}
    \label{table:sig}
\end{table}

\subsection{The Septentrio Mosaic X-5 Receiver}

\begin{figure}[h!]
    \centering
    \includegraphics[width=\linewidth]{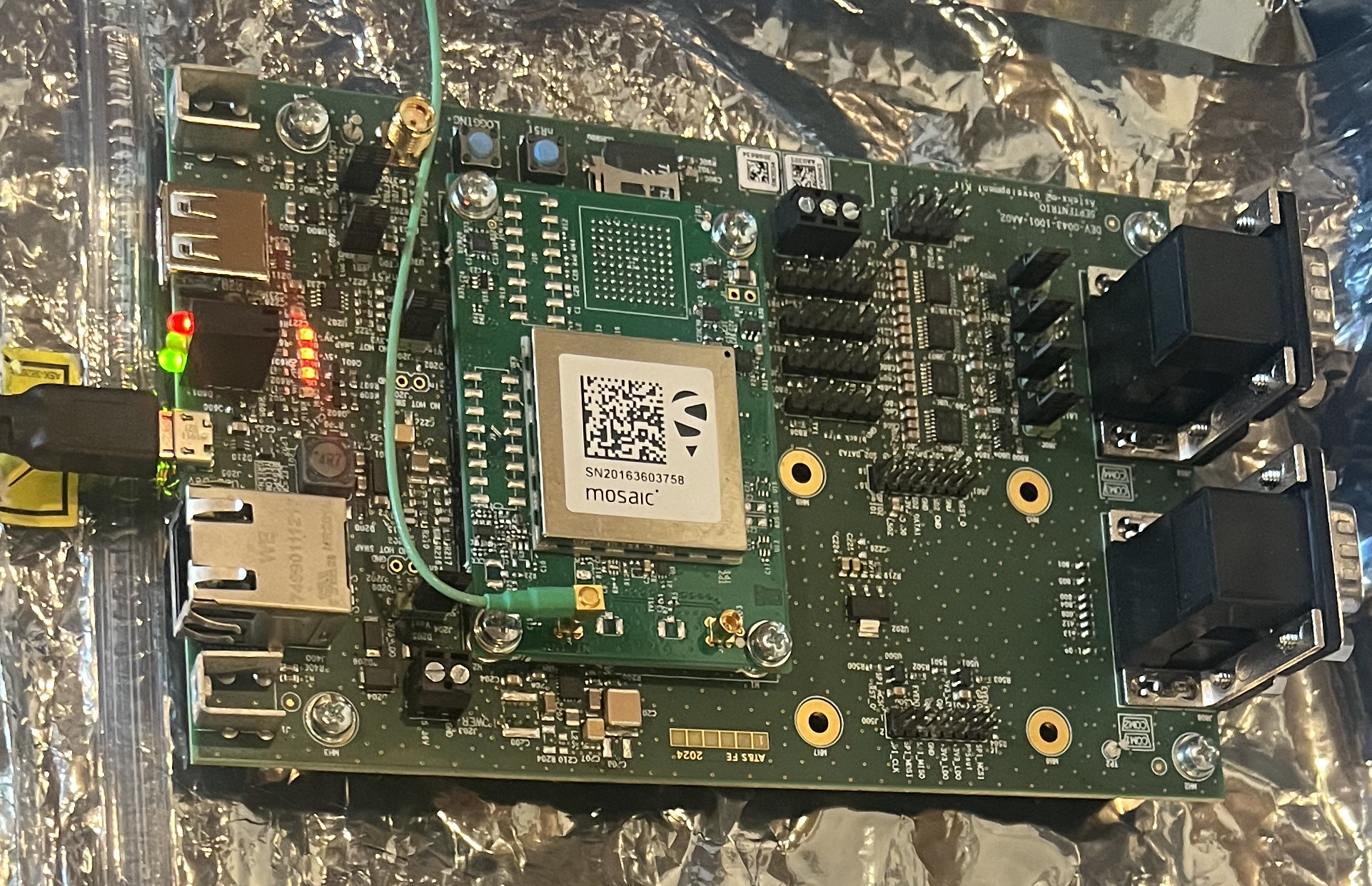}
    \caption{The Septentrio Mosaic X-5 receiver. The receiver can be connected to any computer through a micro-USB port as shown here. The LED lights near the micro-USB port indicate the receiver is on. To check whether the receiver and connected antenna are receiving any satellites, a web server can be opened on a browser on any Windows computer. The RF input comes through the green wire's micro-SMA port. The Mosaic X-5 GNSS module sits on top of the development kit and extracts information from visible GNSS antennas. The receiver was connected to the RF output from D3A.}
    \label{fig:septentrio}
\end{figure}

The Septentrio Mosaic X-5 development kit and receiver are a commercially available GNSS receiver. The Septentrio Mosaic-X5 module is seen in Figure~\ref{fig:septentrio} on top of the development kit itself. The receiver supports multiple connection interfaces, such as micro-USB, Ethernet, and serial ports. We collected data with a PC attached via micro-USB while the receiver is connected to the RF output of the D3A. We inserted a 3 dB attenuator between the D3A output and the receiver input to reduce the signal level. There is also a bandpass filter that acts as a DC block for the 5V usually used to power the active GNSS antenna. The Mosaic X-5 has 448 hardware channels and supports simultaneous tracking of all publicly available and visible satellite signals. It is often updated to include new frequencies and satellites. GNSS receivers work by correlating all potential satellite template signals with the actual received signals at different time delays using matched filtering. This allows the receiver to detect, acquire, and track all visible GNSS satellites even in noisy environments. 

Septentrio reports a horizontal and vertical accuracy of 0.6 cm and 1 cm, respectively, or approximately 30 picoseconds. This does not have an impact on beam mapping but could be important if GNSS is used as a phase/time calibrator. More relevantly, Septentrio lists a resolution for $C/N_0$ measurements of $0.03125~\rm dB-Hz$. They also report less than 45 seconds to a satellite lock after turning the module on, which is consistent with our observations in this study. Septentrio also includes a suite of tools to analyze the GNSS data, namely \href{https://www.septentrio.com/en/products/software/rxtools}{RxTools}, which we use to extract GNSS power, position and satellite names in our own beam mapping analysis. We use the GNSS measurements from the SBF (Septentrio Binary Format) files.



\section{GNSS D3A Beam Mapping Results}
\label{sec:results}
\subsection{Satellite Tracks}
\begin{figure}
\centering
\includegraphics[width=\linewidth]{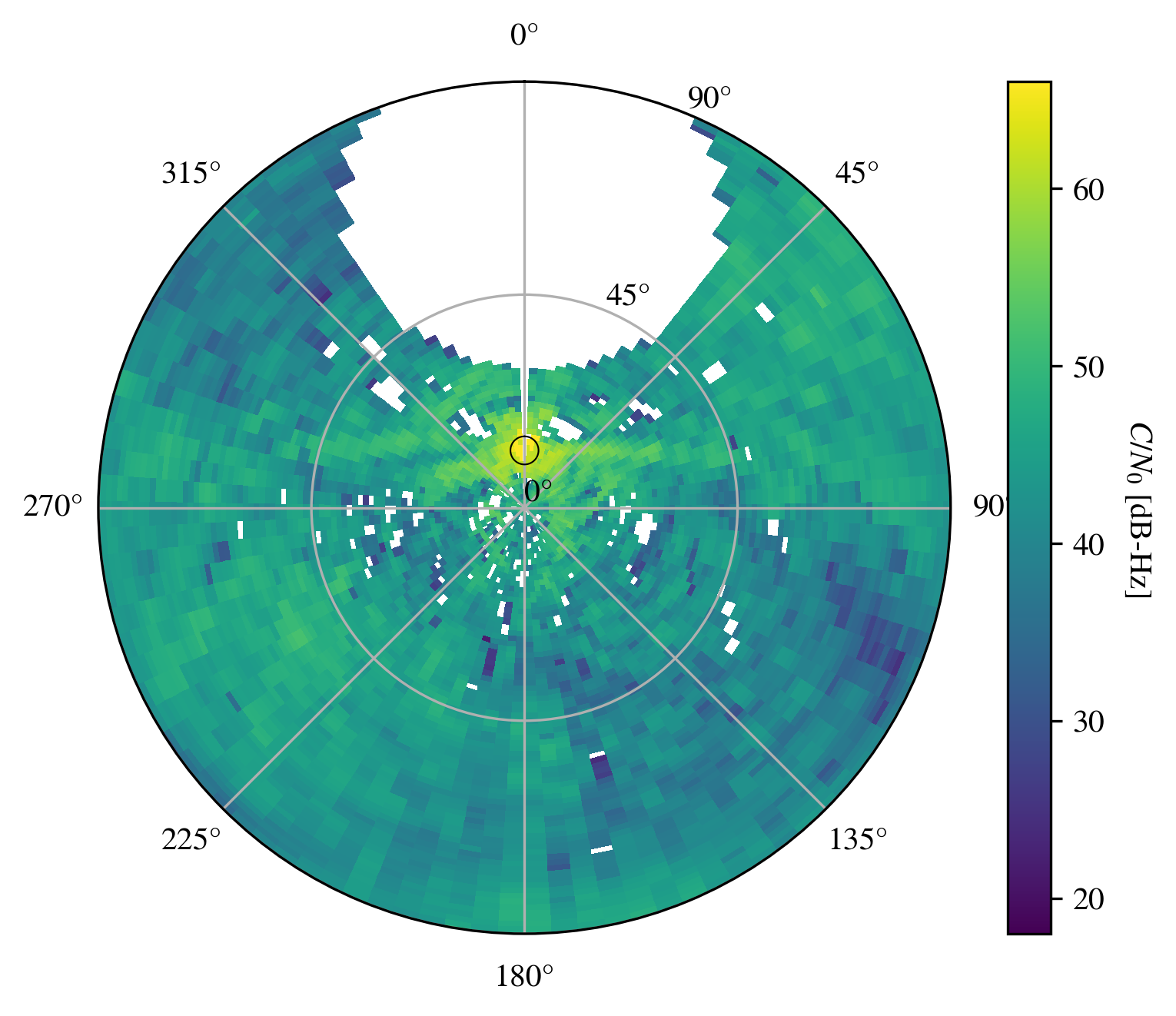}
\caption{Angles are geometric elevation ($0^{\circ}$ zenith, $90^{\circ}$ horizon) and azimuth. The blank region (‘missing satellites hole’) corresponds to sky areas that are not traversed by GNSS satellites due to their orbital trajectories, rather than a true absence of antenna response \citep{Kou03072021}. The tracks of all 85 satellites seen in L1 (for GPS, L1 is at 1575.42\,MHz) visible from the D3A at DRAO over approximately three days between August 30, 2022 at 00:21:57 UTC to September 2, 2022 at 19:25:37 UTC, taken with the Septentrio Mosaic X-5 receiver. We bin the satellite tracks into 10,000 bins in elevation and azimuth in polar coordinates. These include satellites from GPS, GLONASS, and Galileo. The \textbf{maximum} binned power ($C/N_0$) at a particular position is plotted in colour. The beam centre is plotted as a black circle centred at (elevation, azimuth) = ($80.5^{\circ}$, $0^{\circ}$) with a radius equal to the full width at half maximum of the beam at 1.5 GHz. Note that the power plotted here is a \textit{raw} output from the Septentrio receiver and includes both the GNSS beam transmission and D3A power.}
\label{fig:ex_satellites}
\end{figure}

\begin{figure}
    \centering
    \includegraphics[width=\linewidth]{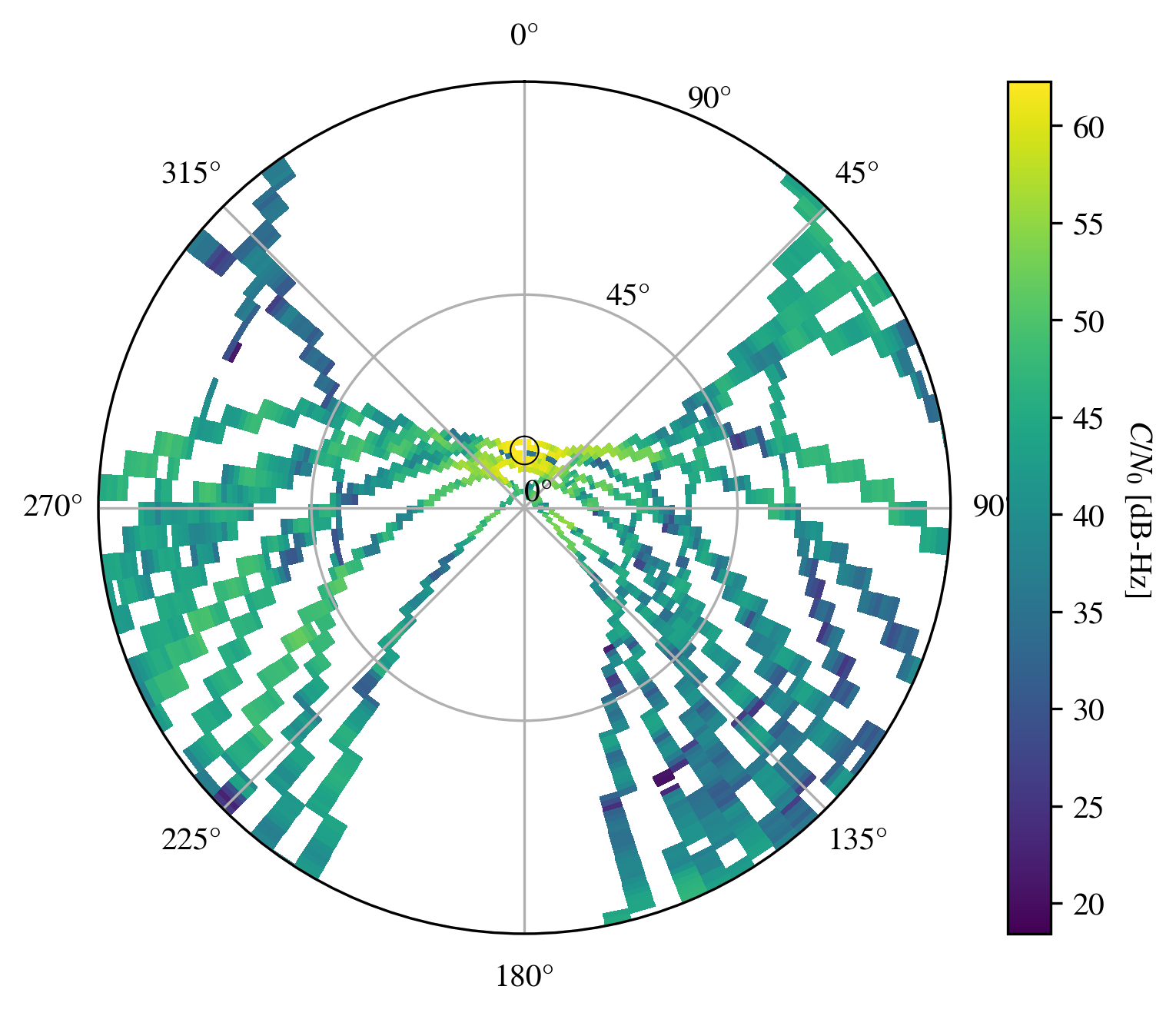}
    \caption{Same as panel~\ref{fig:ex_satellites} but for the tracks of the 6 most repeatable satellites that are shown in 1D in Figures~\ref{fig:time_panel} and \ref{fig:theta_panel}.}
    \label{fig:sub_ex_satellites}
\end{figure}

We used data taken over approximately three days between August 30, 2022 at 00:21:57 UTC to September 2, 2022 at 19:25:37 UTC. The D3A dish was pointed at (elevation, azimuth) = ($80.5^{\circ}$, $0^{\circ}$). We observed more than 80 satellites through one dish of the D3A in the 800-1200\,MHz and 1200-1600\,MHz output bands of the triplexer. We used the L1 frequency band (1575\,MHz). This is due to seeing the most consistent and powerful  measurements at L1 in these preliminary observations. There were 45 satellites which pass within 10 degrees of the centre of the beam. The maximum binned power for tracks of the satellites seen over the course of approximately 72 hours through the D3A are shown in Figure~\ref{fig:ex_satellites}. These included satellites from GPS, GLONASS, and Galileo. In Figure~\ref{fig:sub_ex_satellites}, we select six satellites passing closest to the D3A boresight to clearly see the sharp increase in power.

\subsection{Individual Beam Slices}
\label{sec:indiv}
\begin{figure*}
    \centering
    \includegraphics[width=0.8\linewidth]{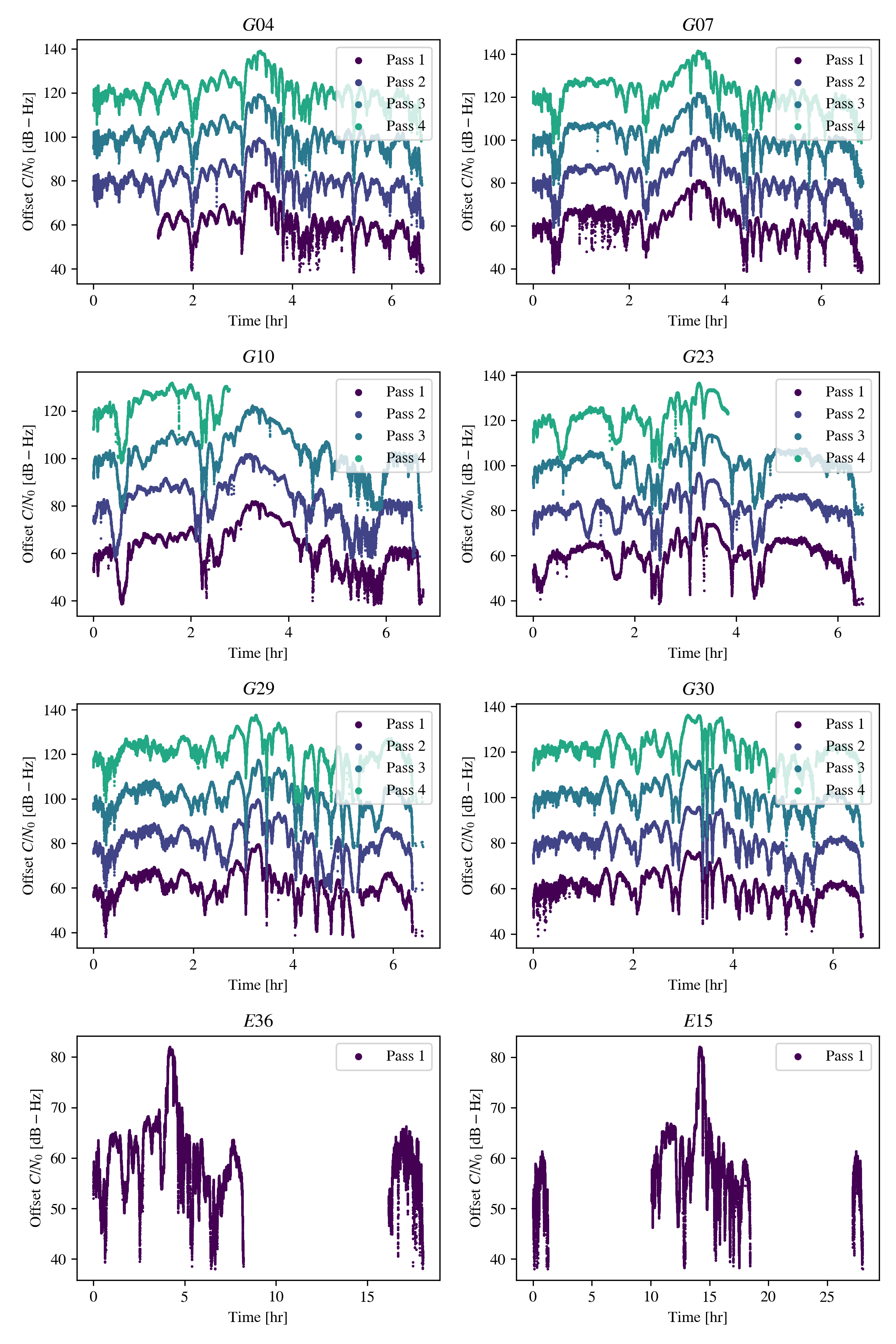}
    \caption{A selection of satellites denoted by their RINEX codes that pass within 5 degrees of the centre of the beam. All observations are of the L1 frequency (1575\,MHz). Each satellite's carrier to noise power ($C/N_0$) is plotted as a function of time since the start of a pass near the D3A. The data were taken over approximately three days between August 30, 2022 at 00:21:57 UTC to September 2, 2022 at 19:25:37 UTC. However, for the European Galileo GNSS satellite with trajectories less suited for North America, we isolate the best pass by eye out of approximately 3 passes. Each pass of the satellite's trajectory is plotted at 20 dB-Hz offset in power. We only include passes of the satellite that are more than 500 measurements long and within the first five passes. Note that the power plotted here is a \textit{raw} output from the Septentrio receiver and includes both the GNSS beam transmission and D3A power.}
    \label{fig:time_panel}
\end{figure*}

\begin{figure*}
    \centering
    \includegraphics[width=0.8\linewidth]{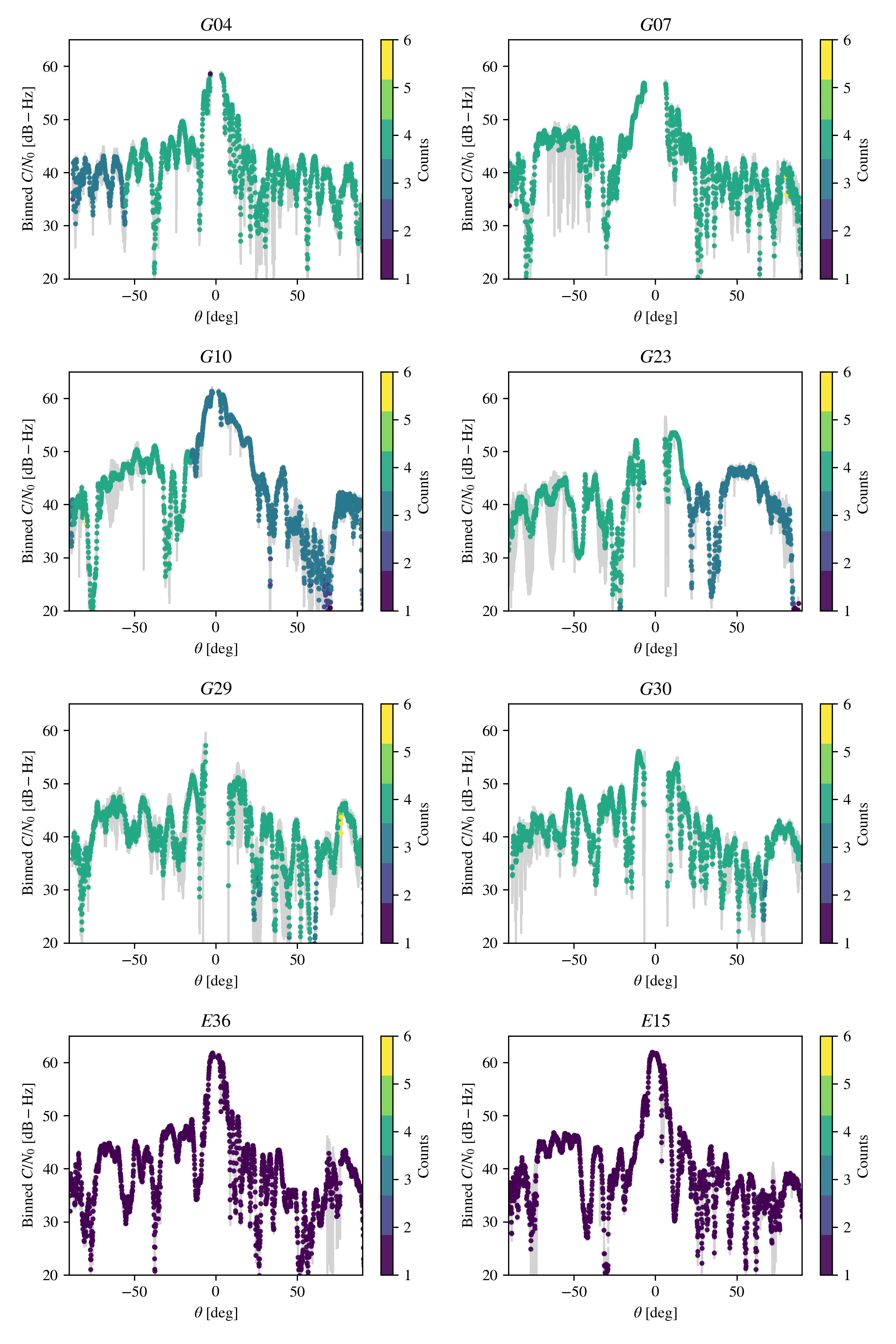}
    \caption{Same satellites as in Figure~\ref{fig:time_panel} with converted averaged $C/N_0$ and angle from boresight. The power measurement is averaged from the measurements shown in Figure~\ref{fig:time_panel} over August 30, 2022 at 00:21:57 UTC to September 2, 2022 at 19:25:37 UTC. Each satellite's carrier to noise power ($C/N_0$) is plotted as a function of angle from the centre of the beam ($\theta$). Following radio astronomy convention, an azimuth of 0 degrees corresponds to due North, with satellites located to the east of the beam centre assigned positive azimuth angles (azimuthal values between 180-360$^{\circ}$) and those to the west assigned negative angles (azimuthal values between 0-180$^{\circ}$). We note that azimuthal variations in passes are not taken into account, and many of the variations outside of the main lobe are due to this. We average the power measurements of each point into widths of 0.1 degree. Thus, the number of counts does not directly relate to the number of passes in Figure~\ref{fig:time_panel}. The error bar is shown in light grey and shows the full range of power measurements made in the bin. The pattern clearly repeats each day as the satellite transits near the centre of the beam at $0^{\circ}$. Note that the power plotted here is a \textit{raw} output from the Septentrio receiver and includes both the GNSS beam transmission and D3A power.}
    \label{fig:theta_panel}
\end{figure*}

When we separate out slices of beam maps made by individual satellites, we see repeatable measurements made over multiple days. We plot each satellite's measured beam power as a function of angle from boresight, $\theta$. We approximate $\theta$ by taking the dot product in Cartesian coordinates between the D3A location and satellite location. To allow for comparison over multiple days, we do not account for the azimuthal variations in the beam for each beam pass and limit our comparisons to the main lobe. This can be seen in Figures \ref{fig:time_panel} (in time) and \ref{fig:theta_panel} (in angle from boresight).  We plot the \textit{raw} output beam data from the Septentrio receiver with minimal processing. The shape of the beam profiles varies as each cuts through a different slice of the beam. Each of the individual satellite beam profiles in Figure~\ref{fig:time_panel} shows the data over the days of acquisition. We offset the beam profile each day to show the repeatability in time. As described in Appendix~\ref{subsec:chunk} and seen for GPS satellite G04 in Figure~\ref{fig:chunking}, satellites do not necessarily pass over the same trajectory each day or were not covered completely in their trajectory. This explains why only part of the G04 primary beam appears in Figure~\ref{fig:theta_panel}. Part of the transit was cut off during data collection. In Figure~\ref{fig:theta_panel}, we convert the beam profiles into an averaged angle from boresight. Since we did not have an external reference GNSS antenna, this means that the D3A beam pattern we plot also includes the GNSS transmission beam. However as seen in Figure~\ref{fig:gnss_beam}, we expect less than a 3 dB difference across the Earth. We do not account for this in these preliminary measurements.

Each beam map shows repeatability in the main lobes that worsens in the sidelobes. Since satellites do not seem to pass through the exact same trajectory each day, we speculate that this is the cause of much of the variability in the sidelobes.  The D3A could be positioned to allow for a GNSS satellite to pass exactly through boresight, but in the sample satellites, we see gaps in the main lobe where the satellites pass close by but not exactly through boresight. 

Despite high variance in some of the sidelobes, which we hypothesize is largely due to azimuthal variations and satellite crowding, we find reasonable agreement between the satellites shown. For the GPS satellites, G23 and G29, we find the smallest difference of 0.56 dB-Hz in the main lobe of the beam. The power is averaged in log space due to not being able to convert to an absolute power measurement. Similarly, the standard deviation is calculated as follows in log space, 
\begin{equation}
    \sigma = \sqrt{\frac{\Sigma_{\rm N} (C/N_0 - \overline{C/N_0})^2}{N}},
\end{equation}
where N is the number of passes the satellite makes through a particular elevation and azimuth. The average number of passes per measurement is 3.
As seen in Table \ref{tab:satellite}, the standard deviation is minimum at the peak of the beam measurement for satellite G04 at 0.09 dB-Hz. This shows promising repeatability for future GNSS beam mapping experiments. 

The mismatch in many of these satellite tracks may be partially due to the satellite nodal precession which occurs daily at a rate of approximately $0.05^{\circ}$. Each day the satellite does not pass through the exact same part of the beam as the day before also adding to the variance. This is described in more detail at the end of Section~\ref{sec:GNSS}. However, we expect the majority of the errors to arise from either too many satellites being present in the field of view or insufficient attenuation before the signal reaches the Septentrio receiver. These factors can lead to signal blending and inaccurate main lobe beam measurements as seen in \citet{2009wska.confE..42O}. Additionally, the absence of a reference antenna limits our ability to perform proper calibration.

\begin{table*}[ht]
\centering
\begin{tabular}{|c|c|c|c|c|}
\hline Satellite &  Max - Min $C/N_0$ [dB-Hz] & Peak $C/N_0$ [dB-Hz] & $\sigma$ at Peak $C/N_0$ [dB-Hz] & Max $\sigma$ [dB-Hz] \\ \hline
G04 & 1.69 & 57.25 & 0.09& 4.97 \\ \hline
G07 & 1.00 & 61.36 & 0.11& 4.28 \\ \hline
G10 & 0.94 & 56.68 & 0.16& 6.19 \\ \hline
G23 & 0.56 & 51.19 & 0.10& 4.80 \\ \hline
G29 & 0.56 & 59.48 & 0.11& 8.46 \\ \hline
G30 & 1.44 & 56.22 & 0.17& 3.02 \\ \hline
\end{tabular}
\caption{We tabulate the minimum $C/N_0$ variation for each satellite, defined as the difference between its maximum and minimum $C/N_0$ across all measured angles. We then show the peak $C/N_0$ measurement for each GPS satellite alongside the corresponding $\sigma$ value at that same angle. Each $\sigma$ value is computed at the exact angle of the observed peak and described in Section~\ref{sec:results}, illustrating the small variance seen within the main lobe of the beam. We also report the maximum $\sigma$ per satellite, representing the highest value measured across \textbf{all} angles from boresight. Galileo satellites (E36 and E15) are excluded from this table due to insufficient pass coverage. \label{tab:satellite}}
\end{table*}

\subsection{Comparison to Sidelobes of Beam Simulations}
\label{subsec:simulations}

\begin{figure}[h!]
    \centering
    \begin{subfigure}{\columnwidth}
        \includegraphics[width=\columnwidth]{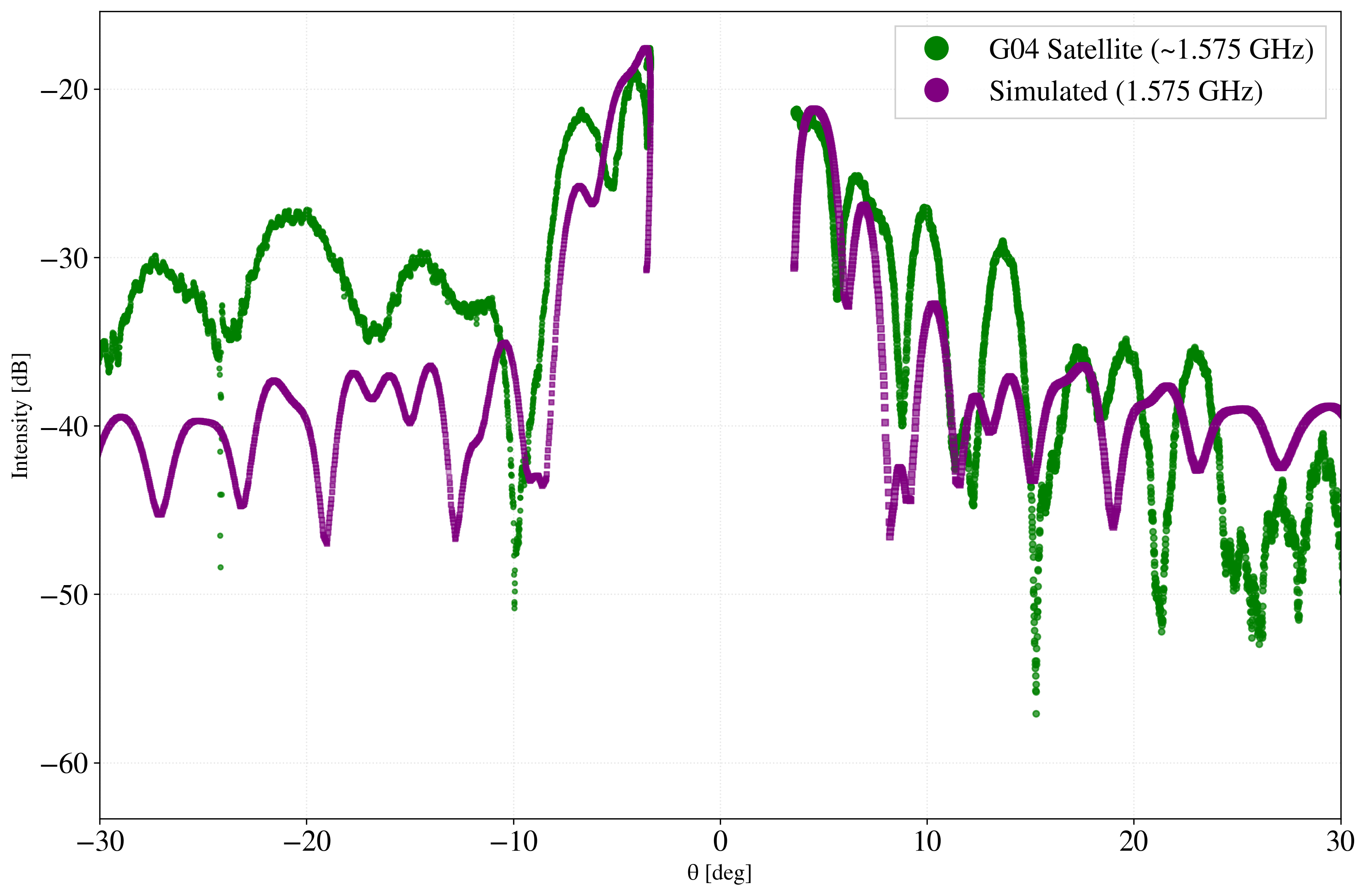}
        \caption{}
        \label{fig:G04}
    \end{subfigure}

    \vspace{0.4cm}

    \begin{subfigure}{\columnwidth}
        \includegraphics[width=\columnwidth]{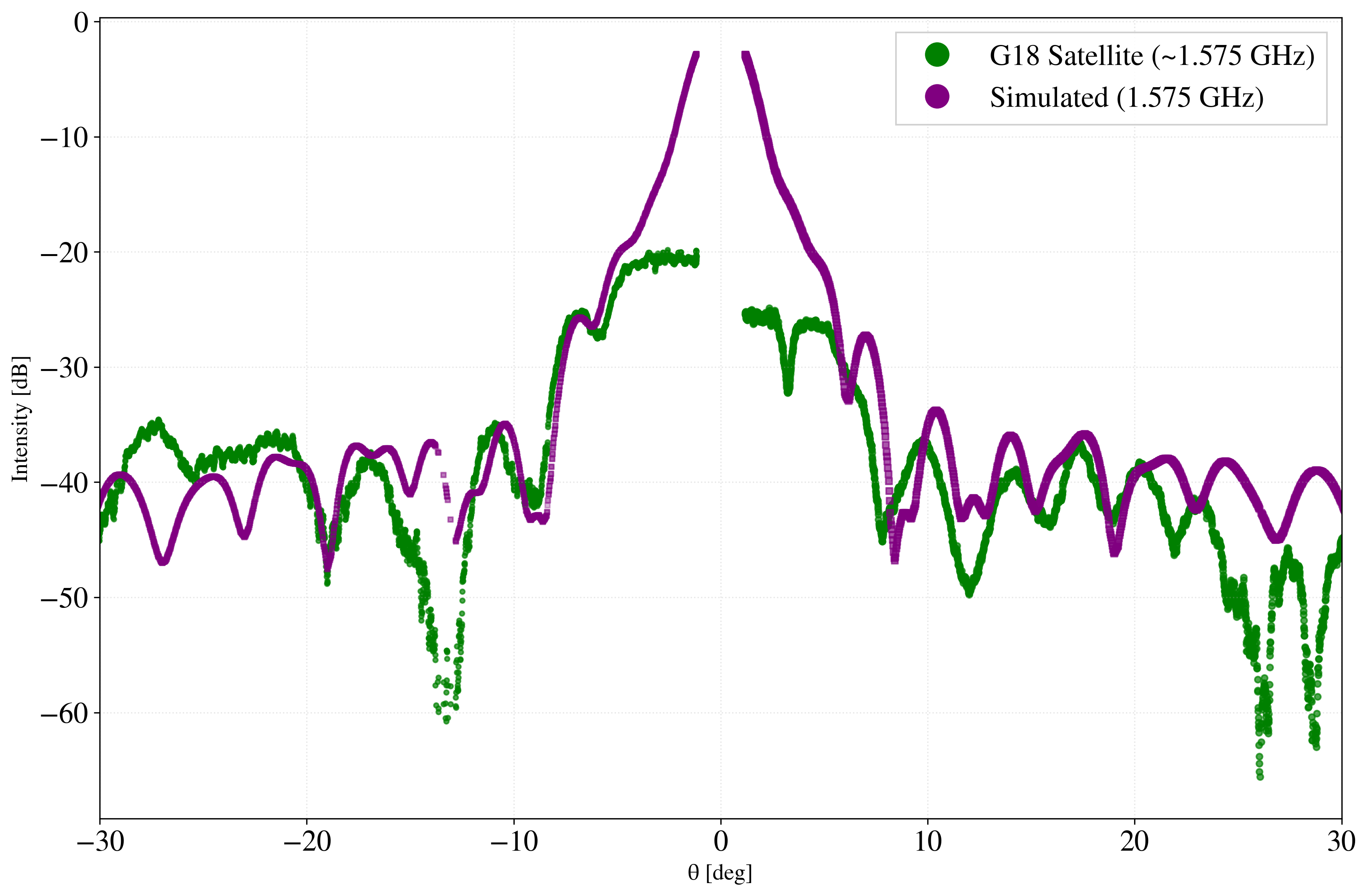}
        \caption{}
        \label{fig:G18}
    \end{subfigure}

    \vspace{0.4cm}

    \begin{subfigure}{\columnwidth}
        \includegraphics[width=\columnwidth]{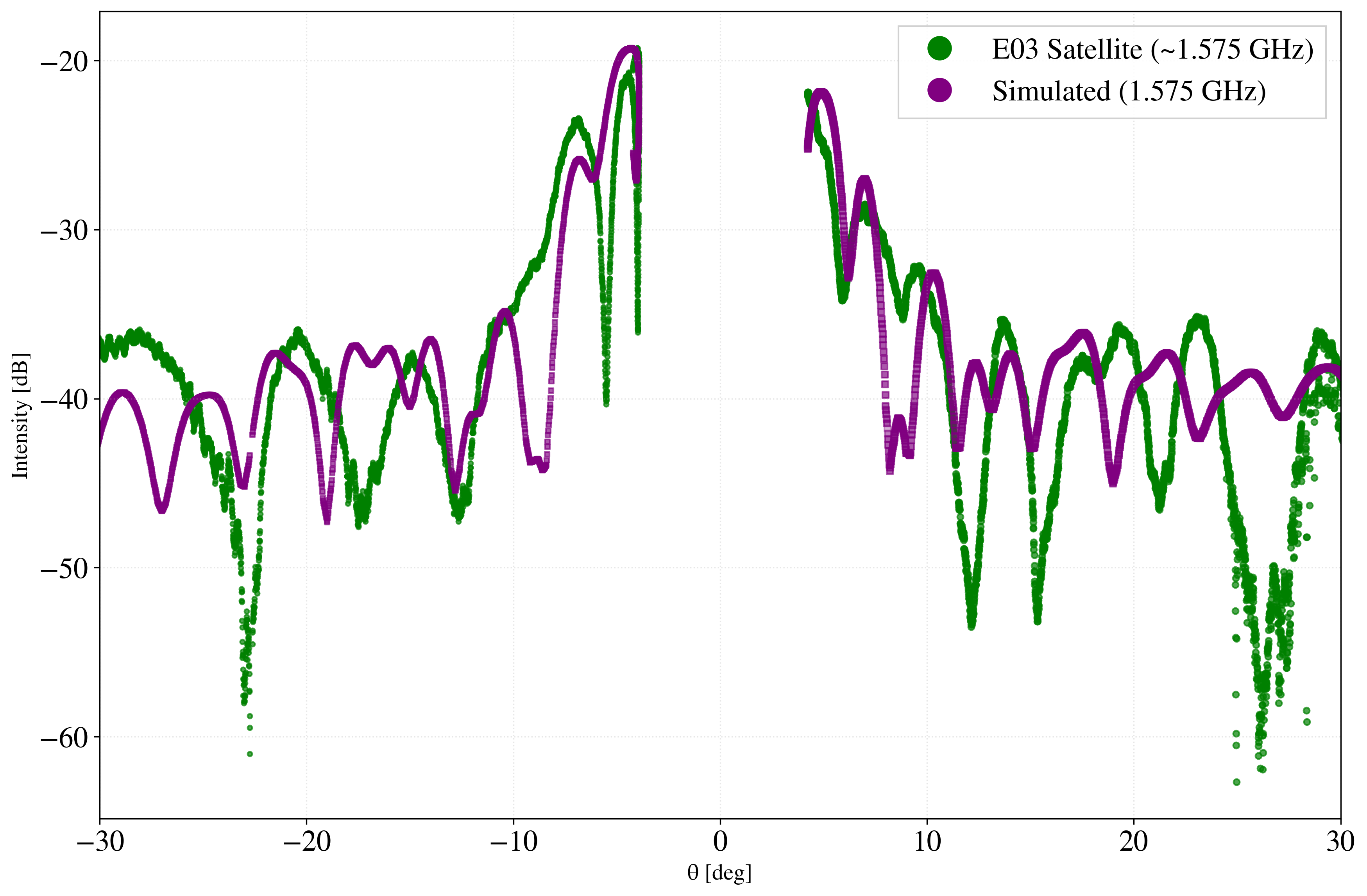}
        \caption{}
        \label{fig:E03}
    \end{subfigure}

    \caption{Comparison between measured and simulated beam patterns for three satellites from the GPS (G) and Galileo constellations (E): (a) G04, (b) G18 (note that we use the first trough, rather than the first peak, to normalise the scaling between this satellite and the simulation), and (c) E03. Some agreement with the simulated beam is visible in certain sidelobes, but discrepancies in the main lobe may result from receiver saturation, track crowding, and limitations in the simulations.}
    \label{fig:bad_sim_panel}
\end{figure}

\begin{figure*}[h!]
    \centering
    \includegraphics[width=0.8\linewidth]{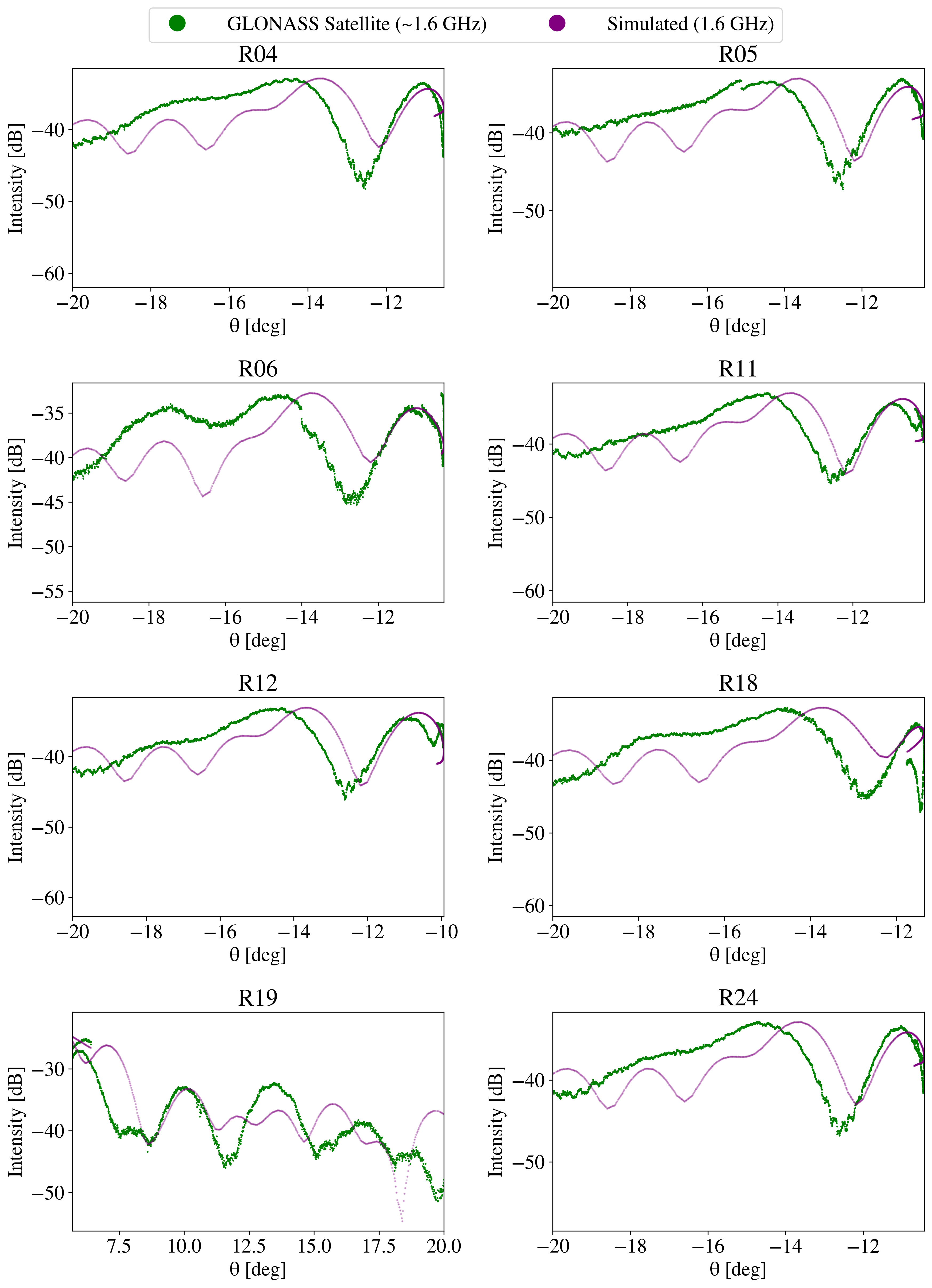}
    \caption{Example GLONASS beam trace showing some agreement with the shape of the simulated beam, particularly in the sidelobes. The clean structure and lack of saturation make GLONASS satellites ideal for characterizing the full beam response at 1.6 GHz. Note that GLONASS satellites do not all emit at exactly 1.6 GHz depending on their satellite number, which likely causes some of the frequency offset seen. Future work is needed to better match the power levels of the simulations and GNSS data (see Section~\ref{sec:discussion}), and may also require improvements to the telescope beam simulations themselves.}
    \label{fig:best_sim_panel}
\end{figure*}

We have shown repeatability in our beam measurements over multiple days. To validate these measurements, we compare them to EM simulations of the D3A performed in CST Studio, as described in Section~\ref{sec:simulated}. Ideally, we would compare the satellite tracks demonstrating repeatability in Figures~\ref{fig:time_panel} and~\ref{fig:theta_panel}. We find that the GLONASS satellite system (as opposed to GPS) most often provides the best match to the simulated sidelobe structure. This is likely due to saturation and satellite crowding in the main lobe for GPS satellites. We show the most repeatable GPS satellite trace (G04) and two others that best match the simulations (G18, E03), along with the single pass GLONASS satellites comparisons, in Figures \ref{fig:bad_sim_panel} and \ref{fig:best_sim_panel}, respectively.

None of the observed satellites follow the same trajectory as celestial calibration sources, precluding direct comparison.  Hence, we use simulations, which can be interpolated to any arbitrary satellite trajectory. We interpolate our 1.575 GHz and 1.6 GHz simulation along the angular paths traced by the satellites, as described in Section~\ref{sec:simulated}. However, GLONASS satellites do not all transmit at exactly 1.6 GHz -- their frequencies vary slightly depending on the satellite number, likely contributing to the observed frequency offset\footnote{See \href{https://gssc.esa.int/navipedia/index.php/GLONASS_Signal_Plan}{here} for the full GLONASS signal plan.}.

Using the angular distance $\theta$ from beam centre—computed from satellite azimuth and elevation relative to the dish pointing—we interpolate the 2D simulated beam map at 1.5\,GHz. Because we lack absolute power calibration, both the measured and simulated traces are normalised by applying a constant offset equal to the difference in their peak intensities (in dB).

Although the detailed intensity profiles do not match perfectly, they exhibit broadly consistent structure out to approximately $-20^\circ$ from boresight. Better agreement is observed at lower intensity levels, as seen in the GLONASS comparison (Figure~\ref{fig:best_sim_panel}) and the sidelobes of G18 (Figure~\ref{fig:bad_sim_panel}) versus G04 and E03, which likely saturate the main lobe (Figure~\ref{fig:bad_sim_panel}). Saturation in the GPS satellites is consistently observed across multiple days (Figure~\ref{fig:time_panel}), but it does not impact repeatability. When scaling the observations and simulations to match in the main lobe power levels, we find deviations within one of the first few nulls of approximately 5 dB or less. This suggests that improved main lobe measurements are feasible, with better agreement potentially achievable in future work through increased signal attenuation during main lobe transits.

\section{Discussion}
\label{sec:discussion}

We summarise and compare existing beam calibration techniques, highlight the unique advantages of GNSS satellites, and outline future steps to develop GNSS into a robust and scalable calibration tool. Among satellite-based methods, GNSS presents a particularly promising option by offering a frequency range nearly 50 times broader than ORBCOMM (which transmits only between 137–150\,MHz), continuous coverage, and far-field conditions suitable for large interferometers.
\subsection{Comparison to Artificial and Astrophysical Calibrators}
Several current and next-generation radio interferometers require precise beam calibration at L-band frequencies (1.1–1.6 GHz). In this regime, GNSS satellites provide a powerful and complementary alternative to more established calibrators. Below, we summarise the strengths and limitations of each approach:

\begin{enumerate}
\item \textbf{Astrophysical calibrators:} While astrophysical calibrators are ideal far-field sources and perform well for high-gain instruments such as the VLA, their limited sky coverage makes them unsuitable for fixed azimuth, transit-style arrays like CHORD. Because they retrace the same sky path each day, they cannot provide the full sky sampling required for beam calibration across all directions. In contrast, GNSS satellites continuously traverse the sky, filling in regions that astrophysical sources never reach (though their orbits still avoid the far north). Separately, the high time cadence of GNSS sources also makes them valuable for transient science, providing frequent and distributed coverage that ensures a calibrator is almost always available. This is not possible with bright pulsars or solar transits, which are too infrequent for daily use.

\item \textbf{ORBCOMM satellites:} Prior work using ORBCOMM (137–150\,MHz) has demonstrated its utility for low frequency arrays such as the MWA and SKA-low. We provide an overview of the potential for the two satellite systems to act as artificial calibrators in Table \ref{tab:freq_coverage_calibration}. For instance, \citet{Line_2018} and \citet{Chokshi_2021} used ORBCOMM to perform dual-polarisation beam calibration of the MWA, achieving residuals within a few dB of simulations. However, ORBCOMM’s narrow bandwidth and limited frequency range constrain its applicability to ultra low frequency studies.  Additionally, ORBCOMM satellites broadcast only intermittently, making it impossible to achieve the high repeatability seen in our GNSS based measurements. GNSS satellites are also  more numerous than ORBCOMM (roughly twice as many visible at a time), enabling higher cadence and denser beam sampling. 

\begin{table*}
\centering
\caption{In this table, we outline the frequency coverage and suitability of selected current and upcoming radio interferometers for ORBCOMM and GNSS beam calibration.}
\begin{tabular}{lccccl}
\hline
\textbf{Instrument} & \textbf{Frequency Range} & \textbf{ORBCOMM} ($f_{\rm emit} = 137-150~\rm$)\,MHz & \textbf{GNSS} ($f_{\rm emit} = 1.1-1.6$GHz) \\
\hline
CHORD     & 300--1500\,MHz          & No  & Yes$^{*}$ \\
DSA-2000  & 0.7--2 GHz             & No  & Yes \\
SKA-low   & 50--350 MHz            & Yes   & No \\
SKA-mid   & 350 MHz--15.4 GHz     & No  & Yes \\
\hline
\end{tabular}
\begin{tablenotes}
\small
\item $^{*}$ The GPS L1 (1.575 GHz) is near the upper end of CHORD’s frequency range.
\end{tablenotes}
\label{tab:freq_coverage_calibration}
\end{table*}

\item \textbf{Drones:} Drones provide high SNR and flexible, controlled source injection, making them highly effective for beam mapping of small dishes. For instance, \citet{Kuhn_2025} mapped the TONE Radio Dish Array \citep{Sanghavi_2024} beam with better than 10$\%$ statistical accuracy using a drone flown at 177 m altitude. While this approach is excellent for individual compact apertures, the limited flight altitude does not satisfy the far-field condition for larger interferometric arrays. In contrast, GNSS satellites at $\sim$20,000 km do satisfy the far-field condition for baselines up to ~45 km at 1.5GHz. While GNSS cannot serve as a true far-field calibrator across the full diameter of very long baseline interferometers, it could still act as an effective far-field calibrator for individual dishes, subarrays, or phased array components. This can offer valuable insight at intermediate stages of the full instrument. 

\end{enumerate}

\subsection{Advancing GNSS Beam Calibration}

GNSS could become a scalable, high cadence beam calibration method in the L-band (1.1--1.6 GHz), a critical frequency regime for next-generation radio telescopes such as SKA-mid and DSA-2000. Our results demonstrate that GNSS satellites can deliver repeatable, daily beam measurements. As seen in Section~\ref{sec:results}, this includes consistent sidelobe detection and reasonable agreement with the shape of EM simulations for CHORD prototype antennas. While this marks an important step forward, further development is needed to fully validate GNSS as a beam calibration method that can be used to advance precision radio astronomy. We outline a recommended path forward to make this a robust beam calibration technique.

\begin{itemize}
\item \textbf{Geometric accuracy} – Precise beam mapping depends on accurate satellite ephemerides and dish pointing. Incorporating high-precision Two-Line Element (TLE) sets will reduce geometric projection and orbital precession errors.
\item \textbf{Absolute power calibration} – These results rely on relative power traces to characterise beam shape. Future work should implement absolute calibration techniques with an auxiliary antenna to enable better absolute power comparisons with EM models.
\item \textbf{Sidelobe confusion and crowding} – The high density of GNSS satellites can lead to overlapping tracks, especially in sidelobes. Dynamic scheduling or filtering for isolated passes can improve data quality.
\item \textbf{Main lobe saturation} – GPS satellites are often too bright, saturating receivers in the main lobe. Selective attenuation or the use of dimmer alternatives (e.g., GLONASS) can mitigate this.
\end{itemize}

\section{Conclusions}
\label{sec:conclusion}

Accurate beam characterisation remains a critical bottleneck in advancing precision radio cosmology, particularly for 21\,cm intensity mapping, where the target signals are $10^{5}$–$10^{6}$ times fainter than the noise. For CHORD, differential measurements between individual element beams are especially crucial. While traditional calibrators such as bright point sources, pulsars, and artificial sources (e.g., drones and satellites) remain useful, GNSS satellites offer a complementary and potentially powerful avenue for beam calibration in the L-band. Ultimately, hardware developments such as integrating GNSS channels into digital correlators may enable lossless recovery of beam information in tandem with science data such as low-$z$ HI galaxies observations. Below, we summarise our conclusions and recommendations for future work.

\begin{itemize}
\item \textbf{Repeatability in GNSS gain measurements} – GNSS shows significant promise as a beam calibration technique between $\sim$1.1–1.6 GHz. We observe high repeatability in beam maps from multiple satellites across four days of data, including measurements into the far sidelobes. Importantly, repeatability between different elements is the most crucial factor for CHORD, making this an important step toward differential beam calibration. Absolute beam power calibration will require an auxiliary antenna to account for satellite gain structure. Crowding and main lobe saturation are limitations but may be mitigated by optimizing observation windows when fewer satellites cross boresight.

\item \textbf{Sidelobes match the shape of simulations} – The shape of GNSS derived sidelobes, particularly from dimmer GLONASS satellites, often shows rough agreement with simulations. We find deviations in one of the first few nulls of less than about 5 dB when matched in power in the main lobe. This highlights the utility of GNSS for probing beam response at large angles from boresight, a regime where other methods struggle. Discrepancies near the main lobe are likely due to saturation and the absence of an auxiliary gain reference, though limitations in the simulations themselves may also contribute.

\item \textbf{Compatibility with next-generation arrays} – Instruments like SKA-mid, CHORD, and DSA-2000 lack reliable far-field astrophysical calibrators across the L-band. With further development, GNSS presents a compelling solution, offering a continuous and global calibration source with a frequency range $\sim$50$\times$ wider than systems like ORBCOMM.

\item \textbf{Recommendations for further development of GNSS beam calibration} – To make GNSS fully viable for use in cosmology and astrophysics radio beam calibration, we recommend the following: (i) deploy auxiliary antennas to enable absolute gain calibration and determine the proper attenuation needed in the signal chain to prevent satellite saturation, (ii) develop pipelines to measure the baseband GNSS signal and preserve spectral and phase information, (iii) model and correct for geometric errors and orbital precession, and (iv) integrate GNSS channels into science correlators to allow for simultaneous calibration and observation. These steps will likely allow GNSS based calibration to meet the precision demands of modern and future radio observatories.
\end{itemize}

Continued development of GNSS as a calibration tool offers a promising path toward precise, wide-angle beam mapping in the L-band. This capability could help enable the next generation of cosmological and time-domain discoveries.
\begin{acknowledgement}
We acknowledge that the DRAO is located on the traditional,
ancestral territory of the Syilx/Okanagan people. Sovereignty was never ceded. \newline \\
The authors thank the anonymous referee for providing constructive comments that improved both the quality of this work and the development of the technique. SB acknowledges Dillon Dong for his expertise and support in the development of this project, which were instrumental in the acceptance of this paper. SB also thanks Pranav Sanghavi, Ian Hendricksen, and Jack Line for valuable discussions on the presentation of the initial beam measurements. Additional thanks go to Matt Dexter, Dan Werthimer, and Quang Tran for helpful conversations and for providing lab space and equipment that supported the completion of this work. SB is supported by the Melbourne Research Scholarship and N D Goldsworthy Scholarship for
Physics. This research made use of Python packages pyuvdata \citep{2017JOSS....2..140H, 2025JOSS...10.7482K}, NumPy \citep{harris2020array}, Matplotlib \citep{Hunter:2007}, AstroPy \citep{astropy:2013, astropy:2018, astropy:2022}, and SciPy \citep{2020SciPy-NMeth}.  
\end{acknowledgement}
\paragraph{Competing Interests}
None
\paragraph{Data Availability Statement}
The code used in the data analysis can be found on GitHub here: \\ \url{https://github.com/sabrinastronomy/mitiono}. RxTools, Septentrio's suite of data analysis and acquisition tools, can be found here: \url{https://www.septentrio.com/en/products/software/rxtools}.\printendnotes\printbibliography\appendix
\renewcommand{\thefigure}{\Alph{section}\arabic{figure}}
\setcounter{figure}{0}
\section{Chunking GNSS Satellite Passes from the Receiver}
\label{subsec:chunk}
\begin{figure*}
\centering
\includegraphics[width=0.8\linewidth]{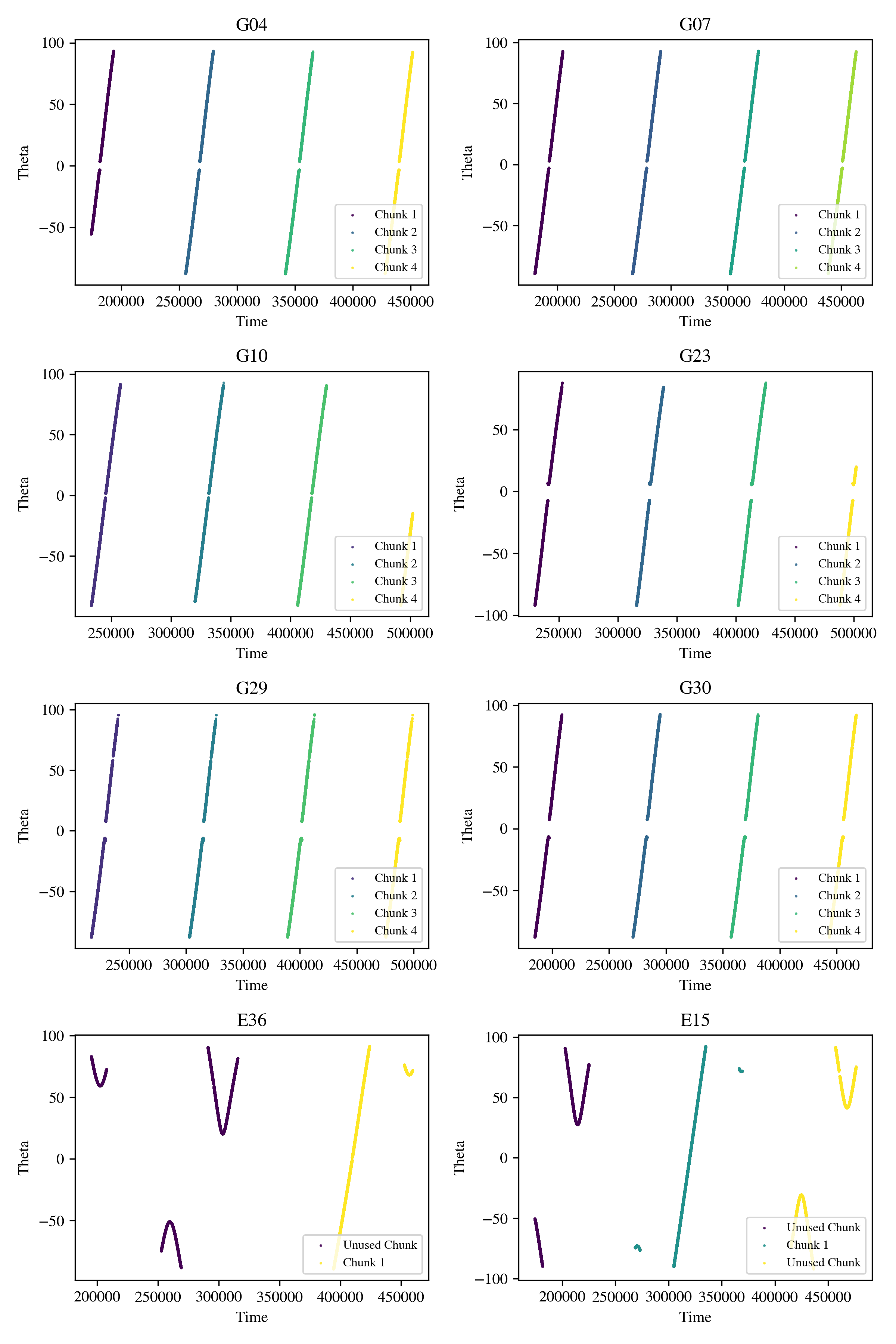}
\caption{Colour-coded chunks of GNSS satellite pass data for all main satellites explored for repeatability in \ref{subsec:chunk} and shown in Figures \ref{fig:time_panel} and \ref{fig:theta_panel}. Gaps in time are used to identify and segment individual passes.}
\label{fig:chunking}
\end{figure*}

To identify individual satellite passes from continuous GNSS observation streams, we implemented a time domain chunking algorithm. Since GNSS satellites periodically rise and set relative to a given observer, large temporal gaps between observations naturally mark transitions between satellite passes. We compute the time difference between consecutive timestamps and define a threshold at $70\%$ of the maximum gap in the dataset. The array is then split at all points exceeding this threshold, effectively segmenting the data into distinct observational chunks. Each resulting segment (or "chunk") contains the associated times, elevation angles ($\theta$), and powers. To visually verify the effectiveness of the chunking, we colour coded each segment and overlaid the resulting chunks in a scatter plot. We display how this is done for the main six satellites explored in this work in Figure~\ref{fig:chunking}. Each colour represents an independent satellite pass.

\end{document}